\documentclass{aa}  

\usepackage{graphicx}
\usepackage{txfonts}
\usepackage{color}
\usepackage{booktabs}
\usepackage{cancel}
\usepackage{amsmath}
\usepackage{soul}
\usepackage{lscape} 
\usepackage{longtable}
\bibpunct{(}{)}{;}{a}{}{,}
\usepackage{ae,aecompl}
\usepackage{color,soul}

\usepackage{subfigure}
\usepackage{booktabs}

\usepackage{float}
\usepackage{tikz}
\usetikzlibrary{shapes,arrows}

\usepackage{natbib}

\usepackage{ulem}
\def\acs606{\textit{F606W}}
\def\acs814{\textit{F814W}}

\def\irac36{${\rm 3.6\mu m}$}
\def\irac45{${\rm 4.5\mu m}$}
\def\irac58{${\rm 5.8\mu m}$}
\def\irac80{${\rm 8.0\mu m}$}
\def\mips24{${\rm 24\mu m}$}
\def\pep100{${\rm 100\mu m}$}
\def\pep160{${\rm 160\mu m}$}
\def\her250{${\rm 250\mu m}$}
\def\her350{${\rm 350\mu m}$}
\def\her500{${\rm 500\mu m}$}
\def\irac{{\rm IRAC\/}}
\def\mips{{\rm MIPS\/}}

\def\otelo{\hbox{OTELO}}

\def\acs{\hbox{HST-ACS}}

\def\pep{\hbox{PEP}}

\def\lephare{\textit{LePhare}}

\def\ha{H$\alpha$}
\def\hb{H$\beta$ }

\def\oiiis{[\ion{O}{iii}]}
\def\oiii{[\ion{O}{iii}]$\lambda$4959,5007}

\begin{document} 

\title{The OTELO survey: Revealing a population of low-luminosity active star-forming galaxies at $z \sim 0.9$}
\author{Rocío Navarro Martínez\inst{1,2},
Ana Mar\'ia P\'erez-Garc\'ia\inst{2,1},
Ricardo P\'erez-Martínez\inst{3,1}, 
Miguel Cerviño\inst{2}, 
Jes\'us Gallego\inst{4}, 
\'Angel Bongiovanni\inst{5,1},
Laia Barrufet\inst{6,7,8}, Jakub Nadolny\inst{9,10},
Bernab\'e Cedr\'es\inst{9,10},
Jordi Cepa\inst{9,10,1},
Emilio Alfaro\inst{11},
H\'ector O. Castañeda \inst{12}\textdagger,
Jos\'e A. de Diego\inst{13,9},
Mauro Gonz\'alez-Otero\inst{9,10},
J. Jes\'us Gonz\'alez\inst{13},
J. Ignacio Gonz\'alez-Serrano\inst{14,1},
Maritza A. Lara-L\'opez\inst{15},
Carmen P. Padilla Torres\inst{16,17,1},
and Miguel S\'anchez-Portal \inst{5,1}}

\institute{\tiny$^1$ Asociación Astrofísica para la Promoción de la Investigación, Instrumentación y su Desarrollo, ASPID, E-38205 La Laguna,
Tenerife, Spain \\
$^2$ Centro de Astrobiología (CSIC/INTA), E-28692, ESAC Campus, Villanueva de la Cañada, Madrid, Spain\\
$^3$ ISDEFE for European Space Astronomy Centre (ESAC)/ESA, P.O. Box 78, E-28690, Villanueva de la Cañada, Madrid, Spain\\
$^4$ Departamento de Física de la Tierra y Astrofísica, Instituto de Física de Partículas y del Cosmos, IPARCOS. Universidad
Complutense de Madrid, E-28040, Madrid, Spain\\
$^5$ Instituto de Radioastronomía Milimétrica (IRAM), Av. Divina Pastora 7, Núcleo Central, E-18012 Granada, Spain\\
$^6$ Geneva Observatory, University of Geneva, Ch. des Maillettes 51, 1290 Versoix, Switzerland\\
$^{7}$ RAL Space, STFC Rutherford Appleton Laboratory, Didcot, Oxfordshire, OX11 0QX, UK\\
$^{8}$ ESAC,European Space Astronomy Center, 28691 Villanueva de la Cañada, Spain\\
$^9$ Instituto de Astrofísica de Canarias (IAC), E-38200 La Laguna, Tenerife, Spain\\
$^{10}$ Departamento de Astrofísica, Universidad de La Laguna (ULL), E-38205 La Laguna, Tenerife, Spain\\
$^{11}$ Instituto de Astrof\'isica de Andaluc\'ia, CSIC, E-18080, Granada, Spain\\
$^{12}$ Departamento de F\'isica, Escuela Superior de F\'isica y Matem\'aticas, Instituto Polit\'ecnico Nacional, Mexico D.F., Mexico\\
$^{13}$ Instituto de Astronom\'ia, Universidad Nacional Aut\'onoma de M\'exico, Apdo. Postal 70-264, 04510 Ciudad de M\'exico, Mexico\\
$^{14}$ Instituto de F\'isica de Cantabria (CSIC-Universidad de Cantabria), E-39005 Santander, Spain \\
$^{15}$ Armagh Observatory and Planetarium, College Hill, Armagh, BT61 DG, UK\\
$^{16}$ INAF, Telescopio Nazionale Galileo, Apartado de Correos 565, E-38700 Santa Cruz de la Palma, Spain\\
$^{17}$Fundación Galileo Galilei - INAF Rambla José Ana Fernández Pérez, 7, 38712 Breña Baja,
Tenerife, Spain}

\date{Received ------ / Accepted -----}

  \abstract
   {}
   {We study a sample of \hb emission line sources at $z\sim$0.9 to  identify  the star-forming galaxies sample and characterise them in terms of line luminosity, stellar mass, star formation rate, and morphology. The final aim is to obtain the \hb luminosity function of the star-forming galaxies at this redshift.
}
  {We used the red tunable filter of the instrument Optical System for Imaging low Resolution Integrated Spectroscopy (OSIRIS) at Gran Telescopio de Canarias (GTC) to obtain the pseudo spectra of emission line sources in the OTELO field. From these pseudo spectra, we identified the objects with \hb emission. 
   As the resolution of the pseudo spectra allowed us to separate \hb from \oiiis, we were able to derive the \hb flux without contamination from its adjacent line. Using data from the extended OTELO catalogue, we discriminated AGNs and studied the star formation rate, the stellar mass, and the morphology of the star-forming galaxies.}
   {We find that our sample is located on the main sequence of star-forming galaxies. The sources are morphologically classified, mostly as disc-like galaxies (76\%), and 90\%  of  the  sample  are  low-mass galaxies ($M_*<10^{10}\;\mathrm{M}_\odot$). The low-mass star-forming galaxies at $z \sim 0.9$ that were detected by OTELO present similar properties as low-mass star-forming galaxies in the local universe, suggesting that these kinds of objects do not have a favorite epoch of formation and star formation enhancement from $z \sim 1$ to now. Our sample of 40 \hb star-forming galaxies include the faintest \hb emitters detected so far. This allows us to constrain the faint end of the luminosity function for the \hb line alone with a minimum luminosity of $\log L = 39 \;\mathrm{erg\,s}^{-1}$, which is a hundred times fainter than previous surveys. The dust-corrected OSIRIS Tunable Emission Line Object survey (OTELO) \hb luminosity function established the faint-end slope as $\alpha =-1.36\pm 0.15$. 
   We increased the scope of the analysis to the bright end by adding ancillary data from the literature, which was not dust-corrected in this case. The obtained slope for this extended luminosity function is $\alpha = -1.43\pm 0.12$.
   }
   {}
\titlerunning{OTELO survey V}
\authorrunning{Rocío Navarro et al.}
\maketitle

\section{Introduction}
The study of star formation along cosmic time is a key tool for characterising the evolution of galaxies and the physics involved at cosmological scales. All the processes concurring in a given space and time affect this activity and, at the same time, are reflected in how the star formation takes place.

The  star formation rate (SFR hereafter) can be estimated though different  indicators, from the X-ray to radio wavelengths  \citep[see the review by][]{Kennicutt1998,Kennicutt2012}, using both continuum and emission lines. The luminosity of the H$\alpha$ emission line is a reliable indicator which scales linearly with the number of ionising photons produced by young massive stars \citep[e.g. ][]{Hayashi2018, Coughlin2018, Matthee2017, Sobral2015, Fujita2003, Gallego1995}. It is also the reddest (hence less affected by internal extinction) and stronger (hence with a lower relative correction due to stellar absorption) Balmer line in the optical, although formally any other Balmer line may also be used  \citep[see][for a discusion about the SFR calibration]{Cervino2016}. For redshifts higher than $z=0.4$,  where H$\alpha$ is not available in the optical range, the second best choice is \hb as the second stronger recombination line.\ We note that there are also some collisional lines which can also be used as a SFR proxy as [\ion{O}{ii}]$\lambda$3727 \citep{Villaverde2010}, although their calibration requires the use of photoionisation codes and it is affected by metallicity and dust attenuation.

Different emission line surveys have been carried out to measure \ha\ and \hb\ spectral lines  \citep[see][among others]{Newman2013, Driver2009, Geach2008, Gallego1995}. These surveys accurately determine the spectroscopic redshifts and therefore constrain the volume under study more accurately. In addition, emission-line surveys allow one to observe fainter galaxies due to the relatively higher brightness of the emission compared to the continuum. Slitless surveys are usually based on  sets of narrow and medium or broad band observations, overcoming the need for a sample of pre-selected objects. However they provide less accurate estimations of redshift and line fluxes. 

In this paper we exploit the data from OSIRIS Tunable Emission Line Object survey (OTELO) (\citealp[hereafter OTELO-I]{Bongiovanni2019}) to analyse the population of \hb\ emitters at $z \sim 0.9$. OTELO is a slitless pencil beam survey that uses the Red Tunable Filter (RTF) of the OSIRIS instrument (\citealt{Cepa98}) at the 10.4m Gran Telescopio  Canarias (GTC, \citealt{Alvarez98}), and it was designed to overcome the drawbacks of integral field surveys by a discrete scanning of the spectral range of interest along the full field of view of the instrument, producing pseudospectra with a resolution of R$\sim$700. OTELO targeted a region relatively free of sky emission lines to find emission lines sources (ELSs) at different co-moving volumes with mean redshifts of up to 6.5. The  RTF was configured to scan a window of 230\AA\ centred at 9175\AA\ in a 7.5 $\times$ 7.4 arcmin$^2$ area located at the south-west edge of the most deeply explored region of the Extended Groth Strip (EGS hereafter). The limiting line flux achieved is 5$\times 10^{-19 }$ erg s$^{-1}$ cm$^{-2}$, which makes it the deepest emission line survey to date (see OTELO-I for a full description of the survey).

The redshift for \hb\ emitters in our sample is centred at $z\sim 0.87$, given the wavelength coverage of the OTELO survey. The spectral resolution of the data makes it possible to deblend \hb\ from its close emision features \oiii{}. This allows us to study the star formation activity at this redshift range, and, using the data in the extended OTELO catalogue, characterise these objects in terms of their photoionisation engine (either active galaxy nuclei (AGNs) or star formation activity), morphology, and stellar mass.

Being able to measure \hb\ without contamination from \oiii\  also allowed us to build the luminosity function (LF hereafter) of \hb\ emitters at this redshift.
Previous works have analysed the LF function with narrow-band images that were not able to resolve the contribution from the \hb line alone, hence obtaining the LF of  H$\beta$+\oiiis\ together. Up to now, only \citet{Comparat2016} have studied a LF of H$\beta$  in the range of $z \sim 0.8$.
However, \citet{Comparat2016}  only cover the bright end of the luminosity function (up to $10^{40}$ erg/s), based on spectroscopic data of VVDS \citep{LeFevre2013} and DEEP2 \citep{Newman2013} surveys. 
We took advantage of the unprecedented depth of OTELO data to extend the \hb\ LF faint end and constrained the number of galaxies at low line luminosities. 

It is well known that the relationship between the SFR and mass for star-forming galaxies (SFGs), the so-called main sequence (MS) of star-forming (SF) galaxies \citep{Speagle2014,Noeske07}. Galaxies on this MS formed stars at much higher rates in the distant universe than they do today. Moreover, the bulk of SF thus appears to have occurred earlier in massive galaxies compared to less massive systems \citep{Bouche10,Elbaz07}. Again, due to the characteristics of the OTELO survey, our sample of \hb\ emitters are low-mass star-forming galaxies. Then, we were  able to analyse the location of low-mass SFGs at $z \sim 0.9$ in the SFR-mass diagram and its impact on the evolution of star-forming galaxies.

This paper is structured as follows. In Section 2 we describe the selection of \hb\ emitters in the OTELO survey and the obtention of the parameters that we use in our analysis. Section 3 addresses the characterisation of \hb\ emitters.
In Section 4 we present the observed LF, including the description of main biases and corrections. The discussion of the results and a comparison with similar data are presented in Section 5. Finally, Section 6 reports the conclusions of this work.

Throughout this paper, we assume a standard $\Lambda$-cold dark matter cosmology with $\Omega_\Lambda = 0.7$, $\Omega_m = 0.3$, and $\mathrm{H}_0 = 70 \,\mathrm{km}\,  \mathrm{s}^{-1}\, \mathrm{Mpc}^{-1}$. All magnitudes are given in the AB system.

\section{The OTELO sample of H$\beta$ emitters}
\subsection{The OTELO catalogue}\label{OTELOdata}

The OTELO-observed window is a region of almost 56 square arcmin in the EGS field centred at RA=14h 17m 33s, Dec=+52$^\circ$ 28$^\prime$ 22$^{\prime \prime}$ (J2000.0), at 36 different wavelengths equally spaced between 9070\AA\ and 9280\AA.\ Using OSIRIS guaranteed time, a total exposure time of 108 hours was dedicated to obtain OTELO data. A source list was extracted from the co-added image, and the flux at each wavelength was obtained and complemented with data from the  CFHTLS survey (T0007 Release), Hubble Space Telescope Advanced Camera for Surveys (HST-ACS), and near-infrared (NIR) data from the WIRcam Deep Survey (WIRDS, Release T0002) to form the core catalogue. Ancillary data from X-ray, ultraviolet (UV), mid-infrared (MIR), and far-infrared (FIR) catalogues were added through educated cross match techniques developed by \citet{PerezMartinez2016}. These data are included in the OTELO multiwavelength catalogue.

The catalogue contains 11237 raw entries with up to 24 photometric detections each. 
The spectral energy distributions (SED) for these were obtained with \lephare\  \citep{Arnouts1999, Ilbert2006}, using a galaxy template library with the four  standard Hubble types \citep{Coleman1980} and six SF galaxy templates  \citep{Kinney1996} to estimate photometric redshifts as described in OTELO-I. 
Filters used to obtain photo-$z$'s cover form 1200 to 10000\AA\  (NUV from GALEX; u, g, r, i, and z from CFHTLS; and J, H, and Ks from WIRDS). The extinction law of Calzetti et al. (2000) was adopted, with values of extinction E(B-V) ranging from 0 to 1.1 in steps of 0.05. The value and quality of the photo-$z$ are included in the final OTELO catalogue. 
The selection of emitting objects candidates is based on the analysis of the pseudo-spectra. The presence of flux excess as part of a possible emission-line  leads to 5322 preliminary emission line candidates. A pseudo-spectrum is the result of the convolution in the wavelength space of the input SED of a given source, with the RTF instrumental response characterised by a succession of airy profiles. 
For more detailed information on the building of the final OTELO catalogue, pseudo--spectra extraction, photometric redshifts estimates, and the selection of emitting objects, readers can refer to OTELO-I and \cite{Bongiovanni2020}. 

\subsection{H$\beta$ emitters selection}
\label{subsec:selection}

In this work we focus on the analysis of a sample of \hb emitting galaxies. To this end, we first describe the implementation of the selection steps used in this
particular science case to identify those that present emission on the \hb line:

Firstly, an initial sample of possible \hb emitters was created by those objects falling in the redshift interval $0.85 < z_{phot}<0.91$ in order to safely include all \hb emission line galaxy candidates. The photo-$z$ solutions that include the OTELO-deep photometry provide 87 preliminary ELSs. From this selection, there were 38 objects with \oiii\  emission, which has been previously studied in \citet{Bongiovanni2020}.

Secondly, each object from this sample of  \hb pre-candidates was analysed using a  web-based interactive graphic user interface (GUI) designed for the analysis of OTELO data\footnote{{\tt http://research.iac.es/projecto/otelo}}. This tool provides the following for each object: the pseudo-spectrum, the $z_{phot}$ solutions obtained by \lephare\ fitting together with the SED of the object and the \lephare\ solutions, the stamps in all broad band filters (including HST and OTELO-deep), and all the available information about the object in the database such as ancillary spectroscopic redshifts provided by DEEP2 of the source if available, and data included in the NASA/IPAC Extragalactic Database (NED), if catalogued. Taking into account all available data for each galaxy, we refined the obtained $z$ by eye verification of the presence of the emission line and the possible $z_{phot}$ solutions, and/or we reject the object as a \hb\ emitter galaxy. After each pre-candidate was analysed by several collaborators following the same criteria, a final sample of true \hb emitters was obtained based on the degree of confidence of the most reliable redshift value assigned by such a process, $z_\mathrm{GUESS}$. We note that such a reliable redshift is usually assigned to the peak of the \hb\ line observed in the pseudo-spectra. From this analysis, we finally identified 47 objects as \hb emitters.


After this process, we obtained 47 objects classified as \hb\ emitters by at least three collaborators. However, six objects are marked as reliable \hb sources, but they present some potential problems as would be a very large uncertainty in the $z_{phot}$ solutions, such as truncated lines (three objects) or possible double sources (three objects). These objects, although classified as \hb emitters, were not included in the analysis and flux measurements, making up a final sample of 41 objects for the analysis. Moreover, we discriminate those sources which are robust bona fide  \hb emitters, meaning they: (i) have a reliable photometric redshift; (ii) have a clearly defined line in the pseudo spectra; and (iii) are well defined in every band. This criteria is met by 28 sources, making up the robust \hb sub-sample. The remaining 13 objects do not fulfill one of the previous terms, but they are also included as \hb emitters.

\subsection{Identifying AGNs}
\label{sec:AGN}

In order to build the luminosity function of \hb emitters at $z\sim0.9$ as well as to infer reliable fluxes, luminosity, and stellar masses, we only used recipes aimed for SFGs.\ Thus, we must discriminate between SFGs and active galactic nuclei (AGNs).

As explained in Section \ref{OTELOdata}, the extended catalogue of OTELO covers from the X--ray to FIR range. First, in order to identify AGNs, we checked if any part of the H$\beta$ sample has emission in soft X-rays, because the high energy emission is the most efficient method to select AGNs. However, there are not X--ray data anywhere in the sample.

On the other hand, we used the IRAC based criteria proposed by \citet{Donley2012}. This method uses the \textit{Spitzer}/IRAC bands (at 3.6, 4.5, 5.8, and 8.0 $\mu$m)   to define an empirical region where the AGNs are located. We found that only one source of our sample satisfied one of these criteria and was hence classified as an AGN galaxy. 
This low fraction of AGNs ($\sim2\%$) can be explained by the fact that OTELO is a very deep survey, but covering a small volume compared with similar surveys ($5190\; \mathrm{Mpc}^{-3}$ for our \hb\ sample, see Section \ref{sec:volume}).
Moreover, we note that only ten objects of our \hb{} emitters have IRAC counterparts ($\sim25$\% of the sample) and, therefore, the fraction of AGNs could be $\sim 8\%$. Finally, due to the low line luminosity of our \hb\ emitters, as we previously noted, we do not expect a noticeable AGN fraction despite the redshift window explored and the AGN fraction depends more on luminosity than redshift \citep{Chiang2019}. Even so,
the obtained fraction of AGNs is consistent with the work of \citet{RamonPerez2019a} who found an AGN fraction of about 7\% for the H$\alpha$ sample of OTELO at $z \sim$0.4. \citet{Bongiorno2010} find a 5\% of type-2 AGNs (from a narrow-line selection) at $z \sim 0.8$ for a sample of 1620 \oiiis\/\hb\ emitters, which is also consistent with our result. 

\subsection{\hb fluxes and luminosities}
\begin{figure*}
    \begin{center}
    \includegraphics[width=0.329\textwidth]{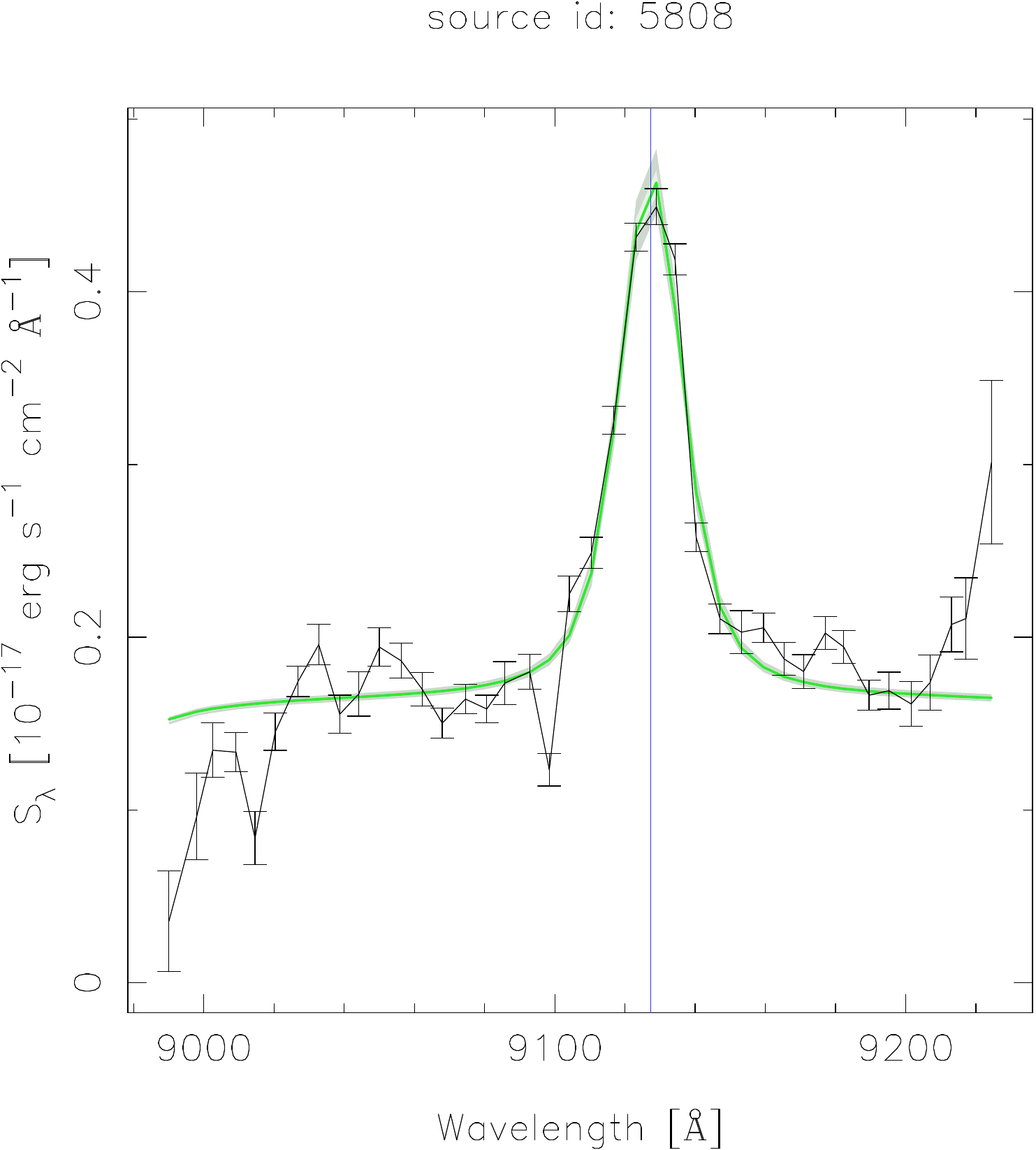}
    \includegraphics[width=0.31\textwidth]{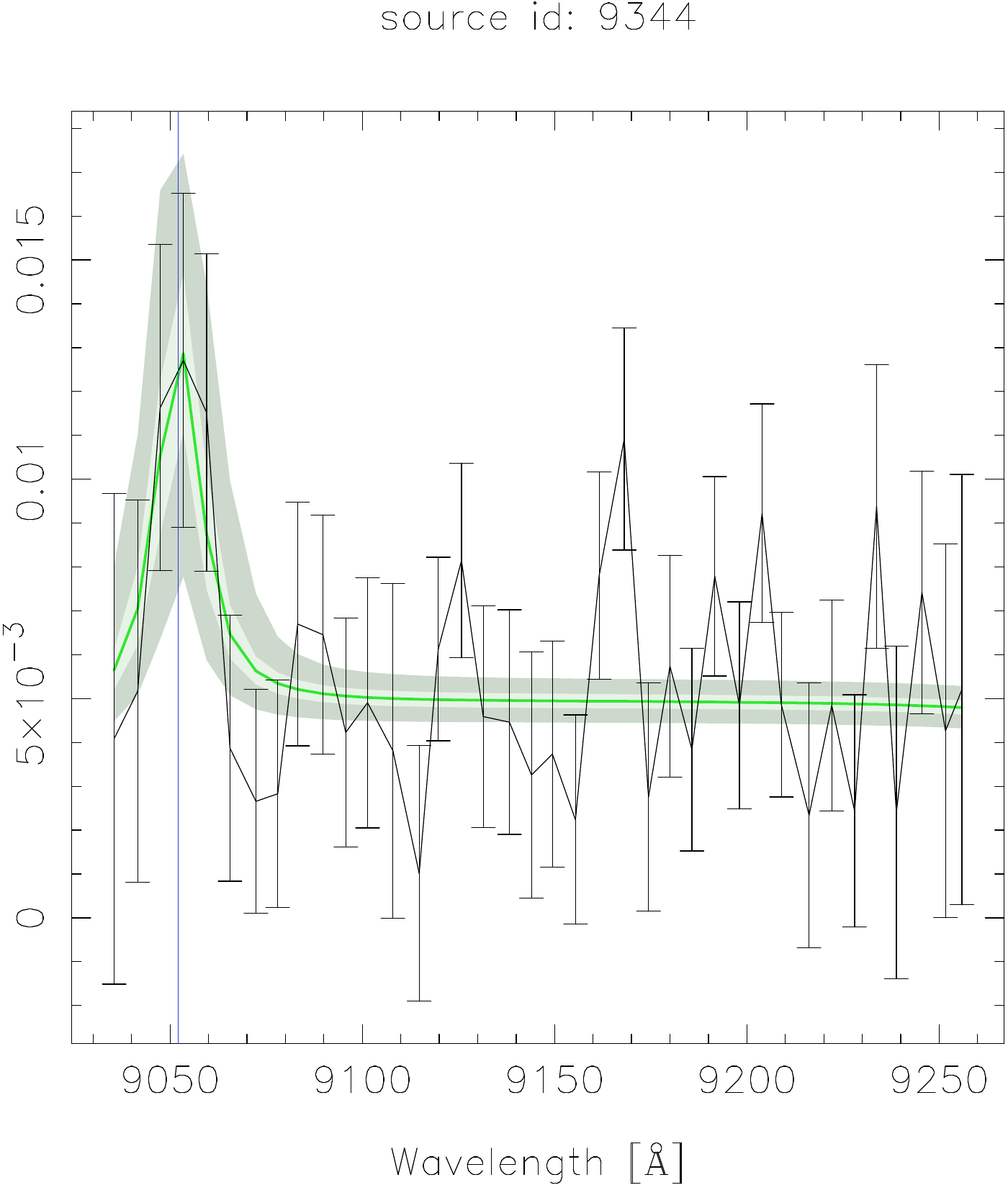}
    \includegraphics[width=0.31\textwidth]{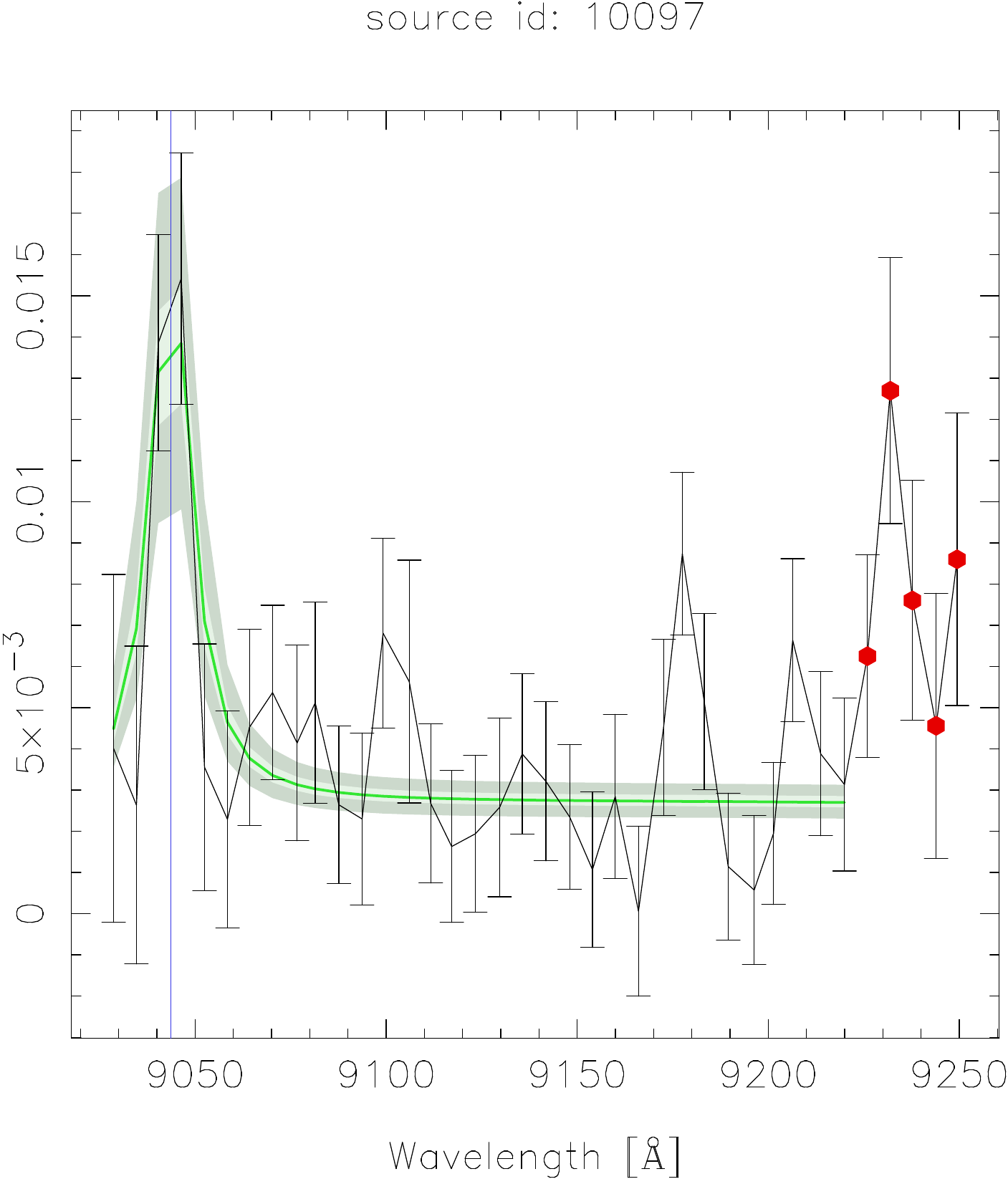}
    \caption{Examples of pseudo-spectra. In each plot, the black line with error bars represents the observed pseudo-spectrum, the green line shows the best fitted result of the deconvolution process, and the light green and light grey shadow area shows the envelope of solutions from the deconvolution process including 25\% and 68\% of the solutions, respectively. The blue vertical line marks the observed \hb wavelength. The left plot ({\tt id:5808}) shows a pseudo-spectrum with a good signal and good fit; the middle plot shows the source {\tt id:9344}, which has a low signal to noise, and therefore a larger uncertainty in the deconvolution results; the right plot shows the case of the source {\tt 10097}, where some spurious points in the pseudo-spectrum were masked during the fitting process.}
    \label{fig:PSexamples}
    \end{center}
\end{figure*}

The pseudo-spectra of the remaining 40 \hb SFG emitters where the flux can be  obtained were analysed in a similar way as described in \cite{Nadolny2020} for the case of $\mathrm{H}\alpha$ sources, based in an inverse deconvolution process of the pseudo spectrum. In summary, we obtain the isophotal flux measured in individual frames as the best approximation for a corrected aperture flux in crowded fields, avoiding then aperture losses. The flux calibration is done using field stars, and it is consistent with the flux density obtained with the SDSS-DR12 photometry (see OTELO-I for details). Then we assume a model spectra as a rest-frame spectra defined by Gaussian profiles of the \hb\ line defined by its amplitude $f_{\mathrm{H}\beta}$ and line width $\sigma$, as well as a constant continuum level of $f_c$ and  $f_\mathrm{mod}(z,f_c,\sigma,f_{\mathrm{H}\beta})$. We performed $10^6$ Monte Carlo simulations where we varied $z,f_c,\sigma$, and $f_{\mathrm{H}\beta}$ in such a way that a likelihood function of all variables was mapped. After that, we marginalised the likelihood function over each of the parameters and obtained the corresponding probability density functions (PDFs). We used the statistical mode as a reference and selected the results within the 68\% confidence interval around this value. For the analysed sample, the resulting PDFs are quite symmetric, allowing, as in this case, the 68\% interval to also be a good proxy for the standard deviation.  All details are similar to those in \cite{Nadolny2020}, except that in our case we sampled the redshift space by following a flat distribution in the range of $z_\mathrm{GUESS} \pm 0.002$ instead of the 0.001 one used in that work.

In Figure \ref{fig:PSexamples} we show three examples of the fit obtained after the deconvolution process. We note that the pseudo-spectra have different qualities and signal-to-noise ratios. Such qualities are consistent with the uncertainties obtained.

\begin{figure}
    \centering
    \subfigure{\includegraphics[width=0.24\textwidth]{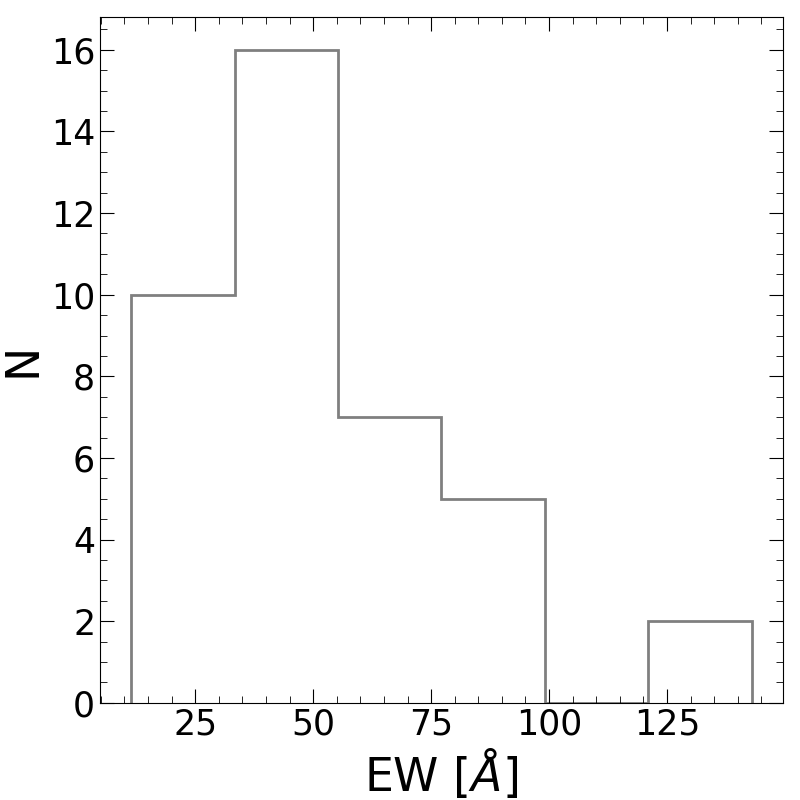}}
     \subfigure{\includegraphics[width=0.24\textwidth]{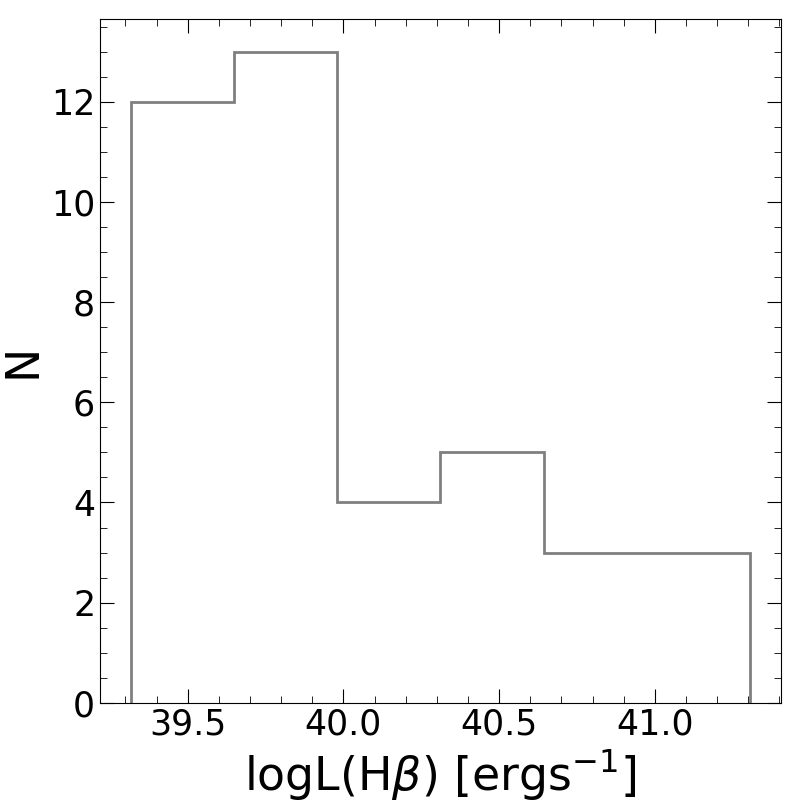}}
    \subfigure{\includegraphics[width=0.24\textwidth]{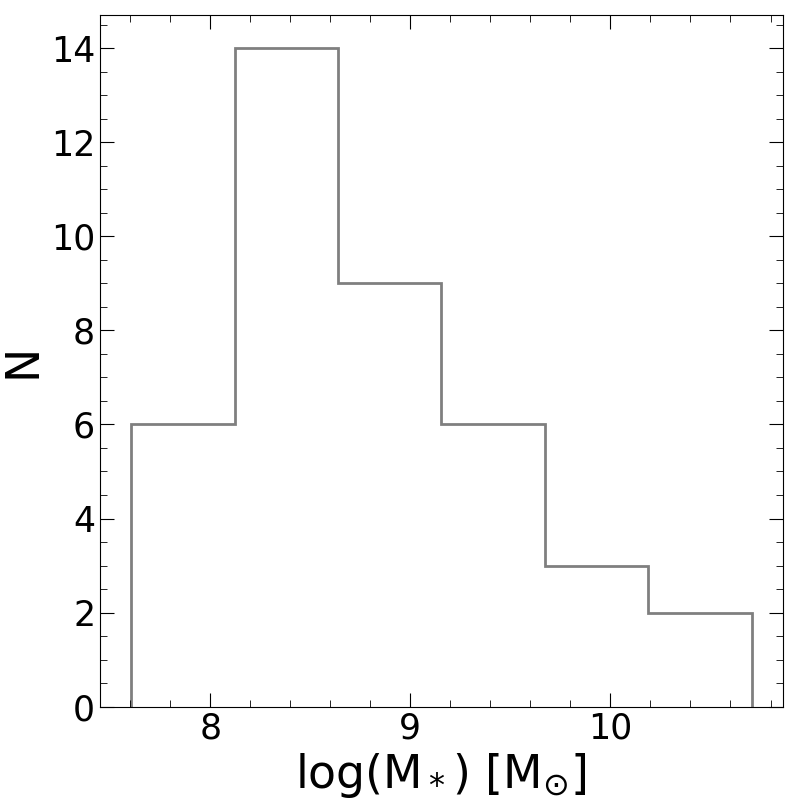}}
    \subfigure{\includegraphics[width=0.24\textwidth]{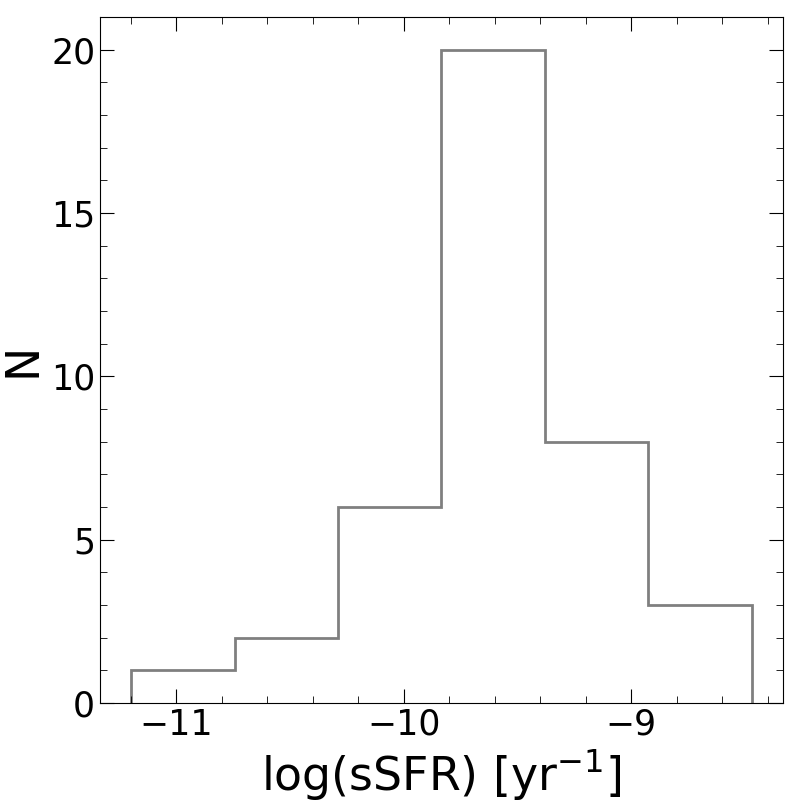}}
    \caption{Distribution of significant parameters  for the H$\beta$ sample.  Top left: Rest-frame equivalent widths. Top right: Corrected luminosity derived from flux. Bottom left: Stellar masses. Bottom right: Specific SFR (see the text for details).}
    \label{fig:histograms}
\end{figure}

It is noteworthy that the \hb\ flux obtained from deconvolution does not take the effect of stellar absorption into account from old stellar populations. Hence, to correct the \hb{} flux of this effect, we adopted the same prescription used by \cite{Nadolny2020}, with a EW=2.5 \AA\ corresponding to the stellar absorption based on \citet{Hopkins2003, Hopkins2013}. We note that such a value is also assumed for H$\alpha$ in \cite{Hopkins2013}. However, these authors show that using the same value for both recombination lines is the best choice based on their simulations. The stellar absorption correction was performed for all sources in a self-consistent way (i.e. over the entire Monte Carlo set and obtaining the resulting PDF and uncertainties). The resulting distribution of EW(H$\beta$) is shown in the top left panel in Fig.~\ref{fig:histograms}. The minimum EW(H$\beta$) is 8.9 with 39 objects with an EW(H$\beta$) larger than 10 \AA, so the choice of the stellar absorption correction has a minor impact on our estimates.

The H$\beta$ fluxes were corrected for extinction by applying the law described in \citealt{Cardelli1989}:

\begin{equation}
    f^{c}_{\mathrm{H}\beta}=f^{o}_{\mathrm{H}\beta}\times 10^{C_{\mathrm{H}\beta}},
\end{equation}

\noindent where $f^{c}_{\mathrm{H}\beta}$ is the extinction-corrected flux, $f^{o}_{\mathrm{H}\beta}$ is the estimated flux from the inverse deconvolution process and corrected for underling stellar absorption,  
$C_{\mathrm{H}\beta}=1.488\times E(B-V)$ is the \hb decrement, and $E(B-V)$ are the broad-band colour excess values obtained from \lephare\ SED fitting,  which includes the intrinsic extinction of the template if it is an SF from \cite{Kinney1996}. The median value of extinction in our sample is 0.1. Also, we looked for MIR and FIR emission in our sources from Spitzer and/or Herschel data, but only a few of the sources, eight to be exact, have detectable emission. These galaxies correspond to the most massive galaxies in the sample ($\log M_* > 9.5),$  with $E(B-V)$ values equal or larger than 0.1. This result is consistent with the inferred low extinction values in the sample.
We are aware that by using the \hb decrement, we assume a common colour excess all along the galaxy independently of its components (stellar and different gas phases) and position.

Finally, the \hb luminosity of each object of our \hb sample was obtained as $L({\mathrm{H}\beta}) = 4\pi f^{c}_{\mathrm{H}\beta}D_L^2$, where $D_L$ is the luminosity distance. The observed luminosity range covered is $39.07 < \log L(\mathrm{H}\beta)[\mathrm{erg}\; \mathrm{s}^{-1}] < 40.98$ (see Figure \ref{fig:histograms}).

\subsection{Stellar masses and SFR}

Stellar masses ($M_*$) for the overall OTELO sample were computed from the mass-to-light ratio prescription of \cite{LopezSanJuan2019} for  star-forming galaxies (more details about obtaining stellar masses in the OTELO survey can be found in \citet{Nadolny2021}. The stellar mass for the \hb sample is in the range of  $10^{7.6} < M_* < 10^{10.7} \;\mathrm{M}_\odot$. 
The resulting distribution of $\log M_*$ is shown in Fig.~\ref{fig:histograms}. We note that the distribution is concentrated at values of $M_*$ below $10^{9.5}$, which points towards a SF nature of the $\mathrm{H}\beta$ emission (see Sect.~\ref{sec:AGN}, instead of AGN hosts).

The SFR of our \hb\ emitters was obtained by following the 
standard calibration of \cite{Kennicutt2012} for solar metalliciy and the initial mass function (IMF) from \cite{Kroupa2001}, but replacing H$\alpha$ by H$\beta$ luminosity:

\begin{equation}
{\rm SFR}({\mathrm{M}_\odot} \,\mathrm{yr}^{-1}) = 5.37\times 10^{-42} \times I({\mathrm H}\alpha)/I({\mathrm H}\beta) \times  L(\mathrm{H}\beta)\,[\mathrm{erg}\; 
\mathrm{s}^{-1}],
\label{eq:sfr}
\end{equation}

\noindent where the factor $I({\rm H}\alpha)/I({\rm H}\beta) = 2.86$ corresponds to the ratio between these lines for solar metallicity Case B recombination and typical electron temperature and density for individual \ion{H}{ii} regions \citep{Storey1995}.
The specific SFR (sSFR) is subsequently estimated as ${\rm SFR}/M_*$. 
The distribution of sSFR is shown in the bottom right panel of Fig.~\ref{fig:histograms} and it is studied in detail in Sect. \ref{subsec:SFR}. The catalogue of the selected sources, including their estimates for $z$, \hb\ flux, EW, stellar masses, SFR, and $g-i$, is shown in the appendix in Tables \ref{tab:emitters1} (the 27 bona fide sources excluding the AGN in the sample) and \ref{tab:emitters2} (the 13 additional sources).

\section{Physical properties of the \hb sample}

As we have described in the previous section, the final sample of \hb emitters that we used consists of 40 objects. This sample of \hb ELSs is distributed in the range from $0.855 \leq z \leq 0.904$. The observed flux is lower than $2\;  \times10^{-16} \, \mathrm{erg}\;\mathrm{s}^{-1}\;\mathrm{cm}^{-1}$ for 83\% of the sample and 73\% of our sources have a rest-frame equivalent width under $60\;$\AA.
Now, we analyse the morphology and colour properties of the sample as a cross-check test of their star-forming nature.


\subsection{Morphology}

To determine the morphological features of our galaxy sample, we made use of the 
available\footnote{\url{http://aegis.ucolick.org/mosaic_page.htm}} HST high-resolution 
(F606W and F814W) images for a visual morphological classification. The detailed morphological analysis of all OTELO sources up to $z=2$ is in the scope of a forthcoming paper \citep{Nadolny2021}.
\\

\begin{figure}
    \centering
     \subfigure{\includegraphics[width=0.24\textwidth]{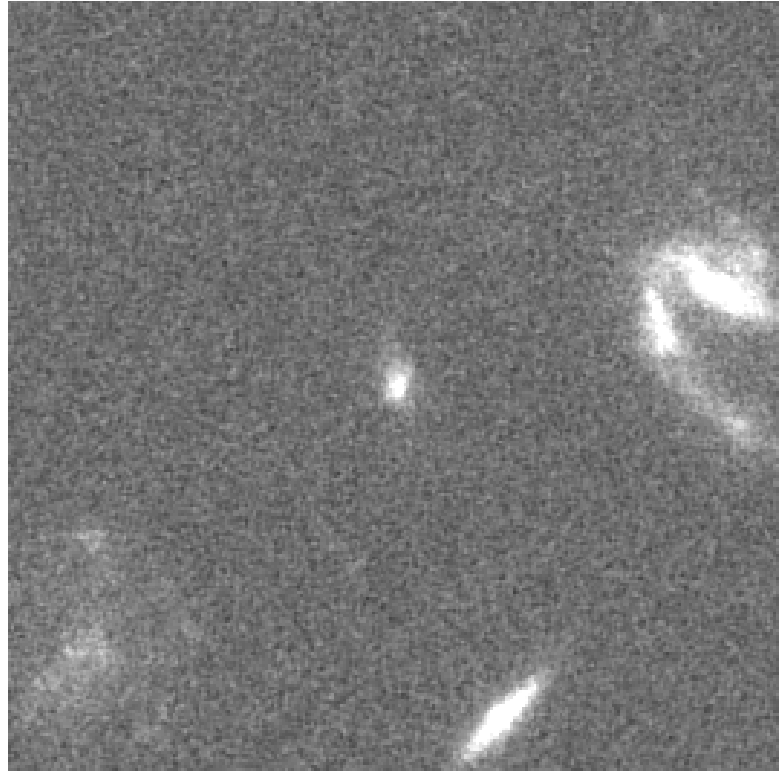}}
    \subfigure{\includegraphics[width=0.24\textwidth]{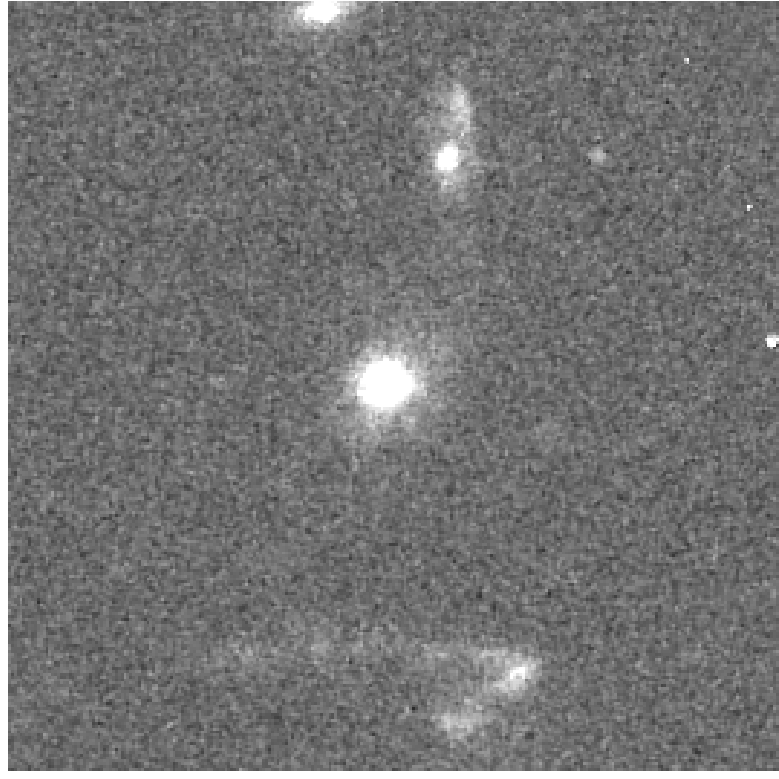}}
    \subfigure{\includegraphics[width=0.24\textwidth]{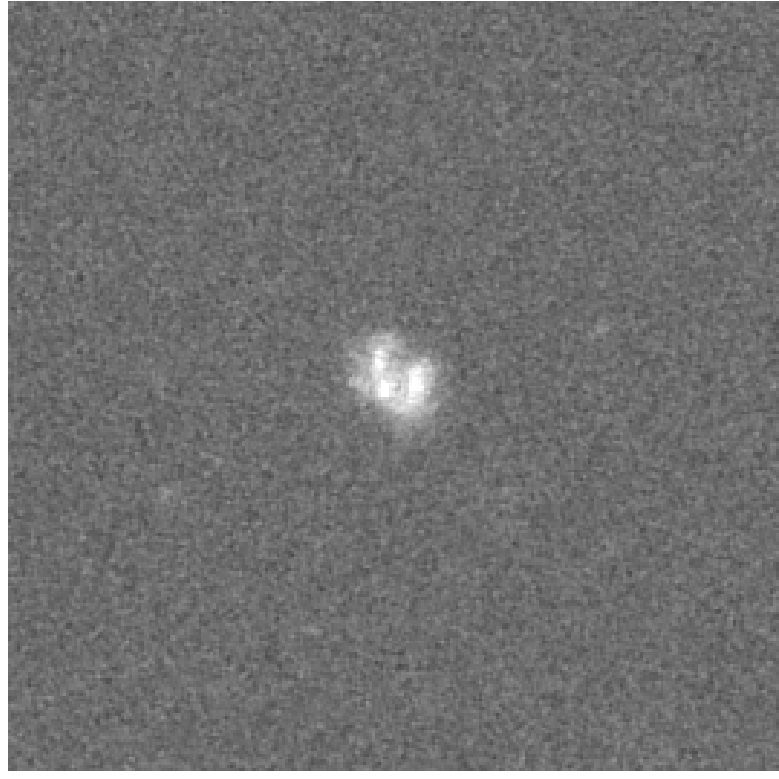}}
    \subfigure{\includegraphics[width=0.24\textwidth]{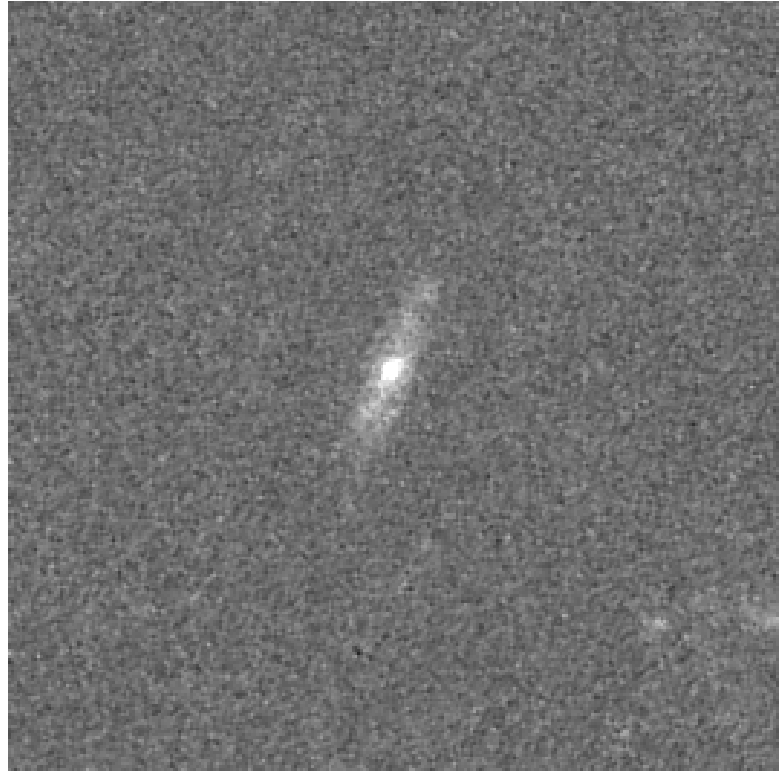}}
    \caption{HST(F814W) images as representative examples of morphology types. The angular size for all of them is 8 $\mathrm{arcsec}^2$. Top left: Disc-like galaxy, which corresponds to the only AGN in our sample. Top right: Elliptical galaxy. Bottom left: Clumpy cluster galaxy. Bottom right: Discy galaxy.}
    \label{fig:morpho}
\end{figure}

Our visual classification is done using MorphGUI, a graphic user interface for morphological analysis developed by CANDELS (see \citealt{Kartaltepe2015}). We modified the interface in order to provide additional morphological classes to extend the classical Hubble scheme to be able to take peculiarities into account (i.e. chain galaxies, tadpoles, clumpy cluster types) found at higher redshifts \citep{Elmegreen2007a}. For 15 sources of the \hb sample, it was not possible to assign a morphology classification. 
Nine of these sources are not detectable in the HST image, and the remaining six  are outside the HST-ACS footprint. The EWs of these 15 objects are larger than 37\AA. Considering that the EW of our sample is above 8.9\AA\, and as stated in \citet{Bongiovanni2019}, the minimum value of EW detected with \textit{p} $\le 0.95$ is 5.7 \AA\ (\oiiis\ at $z=0.8$), we do consider these 15 objects as bona fide sources.
The bulk of classified sources (20, about $\sim$76\%) are  disc-like galaxies, two are early-type, two are clumpy clusters, and one is an interacting system. The morphological classification of the \hb sample shows similar results to those from the OTELO \oiii\ \citep{Bongiovanni2020} sample, where again 85\% of the morphologically classified galaxies are discs, with the HIZELS survey for \ha\ emitters at $z\sim$ 0.8 \citep[$75\pm 8\%$,][]{Sobral2013} and with \citet{Villar2008} in a \ha\ near-infrared narrowband survey at $z=$ 0.84.

Figure \ref{fig:morpho} displays HST images for three different morphology types as well as the AGN host. The sources used for these three morphology examples also have DEEP2 spectra data available, which are included, and the DEEP2 spectroscopic redshift is consistent with our results. We note, however, that the AGN host was excluded from the previous analysis and it is only included here for illustrative purposes.

\subsection{Colour-mass relation}

\begin{figure}
\centering
\includegraphics[width=0.5\textwidth]{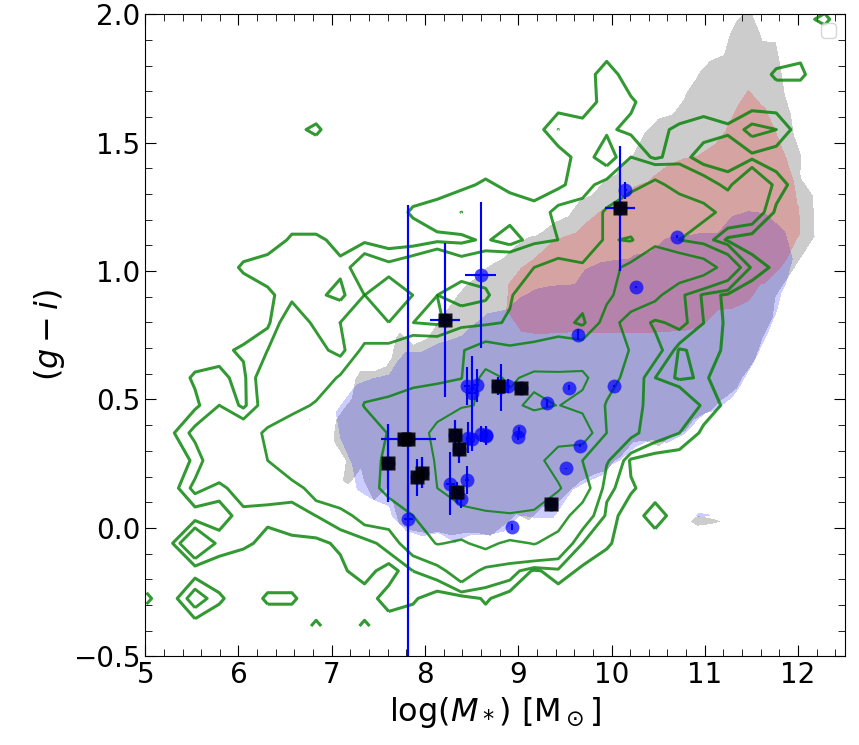}
\caption{Colour--$M_*$ diagram. Blue points show \hb ELS and green contours represent the whole OTELO sample. The filled contour in grey shows the envelope of SDSS-DR7 data, while filled red and blue contours show red and blue clouds separated with the empirically estimated limit of \citet{Bluck2014}. For the \hb sample,  we differentiate the robust sub-sample of 27 galaxies from the remaining 13 with blue circles and black squares, respectively.}
\label{fig:contour}
\end{figure}

Figure \ref{fig:contour} shows the obtained stellar masses as a function of the rest-frame ($g-i$) colour for the star-forming \hb sample. For comparison, in the same
figure, we show the SDSS overall sample together with the blue and red cloud empirical colour division from  \citet{Bluck2014}, obtained using data from SDSS-DR7. Following the same approach as \citet{Nadolny2020}, we identified blue and red clouds using the ($g-r$) colour and represent them in terms of ($g-i$), only for illustrative purposes. Most of the galaxies from the \hb sample are in the blue cloud, as expected, and its stellar masses are mainly in the low mass region ($\mathrm{M}_*<$10$^{10}$M$_\odot$). 

Thus, in summary, our sample is mainly composed of low--mass SF galaxies. These are ideal to study the lower luminosity end of the \hb luminosity function at $z \sim 0.9$ and the low mass end of the SFR properties at such a redshift.

\section{\hb\  luminosity function\label{LFsec}}

{Using our sample of H$\beta$ star-forming galaxies, we obtain the luminosity function, ${\rm LF}({\rm H}\beta$), of OTELO at $z\sim0.9$. The estimation of LF takes completeness and cosmic variance corrections into account, as we describe below. }
\subsection{Survey volume \label{sec:volume}}

The RTF has a characteristic phase effect, so the passband of each pseudo-spectrum is blueshifted  (and slightly narrowed) when the source is further from the optical centre of the image (see OTELO-I for details). Therefore, each source in our sample has been observed at a different redshift depending on its radial distance to the optical centre of the RTF. Taking this phase effect  into account, the range of co-moving volumes is 5130 to 5240 Mpc$^3$. Only as a reference, the characteristic co-moving volume of the sample is 5190 Mpc$^3$ (i.e. the volume corresponding to a radius that separates the field into two equal areas).

\subsection{Completeness correction}

One  of  the  main  challenges  when  deriving  the  LF  is  to estimate  (and  correct)  the  incompleteness  of  the  sample. We performed the simulations of objects' detectability in the pseudo-spectra dominium (flux–continuum and flux–observed line equivalent  width), instead of the common approach of injecting synthetic sources on the real background. In our methodology, a particular simulation is composed of one synthetic pseudo-spectrum for each node of the simulation grid in the Full Width at Half Maximum FWHM-continuum-amplitude space. The synthetic pseudo-spectra are affected by random sky plus photon noise components scaled to the noise distribution of each image (slice)  of the OTELO tomography. Each spectrum was then convolved by the instrumental response of the tunable filter scan to obtain the simulated pseudo-spectra. The parameters were sampled in larger ranges than those covered in the   distributions observed in the real OTELO data. With respect to the noise distribution of pseudospectra, we obtained it by sampling the effective OTELO field in each slice using regions of 1.27 arcsec$^2$. This area corresponds to the mode of the distribution of the effective size of the OTELO sources and it is close to 2 times the area of a point source in OTELO. This methodology is fully described in \citet{RamonPerez2019b} and \citet{Bongiovanni2020}.
This type of data set was fitted by a sigmoid algebraic function (similar in behaviour to e.g. the error function, {\tt erf}) of the form:
\begin{equation}
d = \cfrac{aF}{\sqrt{c+ F^2}},
\end{equation}

\noindent where $F=\log(f_1)+b$, with $f_1$ being the line flux. We assume $a=0.972\pm 0.007$, $b=18.373\pm0.092$, and $c=0.475\pm0.122$, as obtained in \citet{Bongiovanni2020} (see Figure 4 in that work).
This function constitutes the basis of the LF completeness correction.
We note that due to the similarity of the redshifts ranges for \hb\ and \oiiis\ samples in the OTELO survey, we adopted the same function as for \oiiis\ OTELO emitters . 

\subsection{Cosmic variance}
Since the OTELO survey covers a small sky area at about 0.015 deg$^2$, the effects of the cosmic variance (CV) are remarkable \citep{Stroe15,RamonPerez2019b}, especially when compared with surveys with larger volumes (see table \ref{funlumtable}). It is then clear that obtaining an estimation for the CV effects is essential when characterising the \hb luminosity function.

To obtain an uncertainty value due to the CV ($\sigma_{\rm CV}$) for the science case addressed in this work, we firstly tested the approach given in \cite{Bongiovanni2020}, who followed the prescription of \cite{Moster2011}. This approach is based on predictions from cold dark matter theory and the galaxy bias, and it takes into account the surveyed area and the redshift range sampled as input values. The mean CV uncertainty obtained for our \hb sample is close to 0.4. But the mean density of our line emitters at $z = 0.9$ is about $1.2 \pm 0.8 \times 10^{-2}$ Mpc$^{-3}$, which means that the expected CV effects could be even larger and markedly dependant on the \hb luminosity. Hence, we examined the CV estimations using the recipe provided by \cite{somerville04}, which is based on number densities against average redshifts in deep surveys, independently of the clustering strength. We estimated this uncertainty for six luminosity bins. The mean CV value obtained using this estimation is $\sigma_{\rm CV}=0.61$. Noticing the differences between these two estimations, we adopted the \cite{somerville04} prescription since it is more conservative. Table \ref{LFdata} shows the CV estimation for each luminosity bin. As stressed in Section \ref{sec:lumfun}, the uncertainty due to the CV is the greatest contributor to the overall uncertainty of the LF.

\subsection{The luminosity function}
\label{sec:lumfun}

\begin{table}[t]
\begin{center} 
\small
\caption{Binned values of the observed \hb\ luminosity function. Errors in column 2 include all uncertainties described in Section \ref{LFsec}. The fourth column contains the observed number (i.e. before completeness correction) of the \hb\ ELSs in each luminosity bin and the last one shows the cosmic variance parametrisation per bin using the prescription by \cite{somerville04}.}
\addtolength{\tabcolsep}{-1pt}

\begin{tabular}{c c c c c}\\
\hline
\\
$\log L(\mathrm{H}\beta)$ & $\log \phi $ & Number of & Typical &  $\sigma_{\rm CV}$\\ 
$[\mathrm{erg}\;\mathrm{s}^{-1}]$ & $[\mathrm{Mpc}^{-3} \mathrm{dex}^{-1}]$ & \hb\ ELS & Completeness & \\
\\
\hline
\\

39.24 & -2.11$^{+0.21}_{-0.43}$   & 8 & 0.63 & 0.51\\
39.56 & -2.04$^{+0.21}_{-0.41}$   & 12 & 0.78 & 0.54\\
39.88 & -2.14$^{+0.22}_{-0.46}$  & 10  & 0.84 & 0.57\\
40.20 & -2.69$^{+0.26}_{-0.79}$   & 3  & 0.88 & 0.61\\
40.52 & -2.58$^{+0.27}_{-0.85}$ & 4 & 0.91 & 0.70\\
41.84 & -2.71$^{+0.29}_{-1.25}$  & 3 & 0.93 & 0.75\\ 
\\
\hline
\end{tabular}
\label{LFdata}
\end{center}
\end{table}


The \otelo\ survey was mainly designed to obtain a large database of emission-line objects at different epochs. The volume at redshift $z\sim$0.9 stands out in terms of the number of raw ELS candidates obtained, as shown in \otelo\-I (\hb\ and \oiiis\ emitters given the wavelength range covered). 

The survey produced an unprecedented sampling of the faint end of the \hb/\oiiis--LF  due to its low limiting flux, which was obtained by staring at a narrow region of the sky with long exposure times. However, this observational strategy hinders the capability of tracing the bright end, precisely because the small angular size covered ($\sim 0.015\,\mathrm{deg}^2$) decreases the chances of detecting high luminosity galaxies. 

We computed the luminosity for each  galaxy of our ELS sample from the fluxes obtained from inverse deconvolution. The \hb luminosity is distributed in the range $39.07 <\log  L(\mathrm{H}\beta) < 40.98$. Then we computed the number
$\Phi$ of galaxies per unit volume (V) and per unit \hb-luminosity
$\log L$(\hb). This number is provided by: 

\begin{equation}
\Phi[\log L(\mathrm{H}\beta)] = \kappa \cfrac{4\pi}{\Omega} \sum\limits_{i}^{} \cfrac{1}{d_i},
\end{equation}

\noindent where $d_i$ is the detection probability defined above for $i$ galaxies, $\Omega$ is the surveyed solid angle 
($\sim 4.7 \times 10^{-6}$ str), and $\kappa$ is a normalisation factor proportional to $V_{max}^{-1}$, which is the volume 
limited by redshifted \hb\  at the maximal spectral range covered by the \otelo\ scan, including the effect 
of the wavelength variation with the distance to the optical centre mentioned above.\par

The Schechter function \citep{Schechter1976} is the formalism adopted to describe the luminosity function, which is defined as follows:

\begin{equation}
\Phi[\log L(\mathrm{H}\beta)]\ \mathrm{d}\log L = \phi(L) \mathrm{d}L,
\end{equation}

\noindent where $\phi(L) \mathrm{d}L \equiv \phi^\ast (L/L^\ast)^\alpha \exp(-L/L^\ast) \mathrm{d}(L/L^\ast)$. The parameters $L^\ast$, 
$\phi^\ast$, and $\alpha$ are the characteristic value that separates the high and low luminosity regimes in the LF, 
the number density at $L^\ast$,  and the slope of the faint end of the function, respectively.


A Schechter function was fitted to the completeness corrected data given in Table \ref{LFdata}, using a least-squares minimisation algorithm based on the Levenberg-Marquardt method. The uncertainties pertaining to the number density include all the corrections mentioned in this section, including the Poisson error. However, it is worth mentioning that the cosmic variance uncertainty is the main contributor in every bin. The parameters obtained from the completeness-correction LF from OTELO 
are summarised in Table \ref{funlumtable}.

\begin{table*}
\caption[Luminosity functions]{Best-fit Schechter parameters of \otelo\ LF for the \hb\  ELS samples
and its integrals.}
\vspace*{-5mm}
\label{funlumtable}
\begin{center}
\resizebox{\textwidth}{!}{%
\begin{tabular}{c c c c c c c c}
\hline  
Line and & Number & Redshift & ${\rm V}_{c}\tablefootmark{\ (a)}$ & $\rm{log}\phi^*$ & ${\rm logL}^*$ & $\alpha$ & logL range\\
     dataset &of sources & Range & $[{\rm Mpc}^{3}$] & $[{\rm Mpc}^{-3}]$ & $[{\rm erg}\;{\rm s}^{-1}]$ &  & $[{\rm erg}\;{\rm s}^{-1}]$\\

\hline   
\hline
\hb OTELO & 40 & $0.86-0.90$ & 5190 &  -3.08$\pm0.19$ & 41.34(fixed) & -1.36$\pm0.15$ & $39.32-41.31$\\
\hb  OTELO+C\tablefootmark{\ (b)}& 739 & $0.78-0.90$ &  $\sim$10$^{6}$ &  -3.40$^{+0.20}_{-0.23}$ & 41.65$^{+0.11}_{-0.09}$ & -1.43$\pm0.12$ & $39.08-42.5$\\
\hb+\oiiis{} K\tablefootmark{\ (c)}   & 1669 & $0.83-0.85$ &  1.79$\times$10$^{5}$ &  -2.55$^{+0.04}_{-0.03}$ & 41.79$^{+0.03}_{-0.05}$ & -1.6(fixed) & $41.0-42.6$\\
\oiiis{} OTELO   & 184 & $0.78-0.87$ &  6.6$\times$10$^{3}$ &  -2.10$\pm0.11$ & 41.46$\pm0.09$ & -1.03$\pm0.08$ & $39.2-42.0$\\
\hline  
\end{tabular}
}
\end{center}
\tablefoot{}
\tablefoottext{a}{Co-moving volume.}
\tablefoottext{b}{\citet{Comparat2016}}
\tablefoottext{c}{\citet{Khostovan2015}}
\\
\end{table*}

\section{Discussion}

As we have mentioned previously, OTELO is an ultra-deep pencil-beam survey.  OTELO reaches emission-line fluxes as faint as $10^{-19} \mathrm{erg}\,\mathrm{s}^{-1}\,\mathrm{cm}^{-2}$, but it covers a field of view of about 56 arcmin$^2$. This characteristic determines the kind of galaxies that constitute the different populations detected by the survey. The total population is composed of several disconnected emission-line populations at different redshift intervals, selected by the presence of independent emission lines at the corresponding redshift.

In this way, at each redshift we are covering a given volume that is smaller than the typical volumes enclosed by surveys that cover wider apparent fields. The range of the LF that is best traced by OTELO always corresponds to low luminosities, that is, the range of the LF characterised by the exponential slope $\alpha$. This is complementary to regular surveys where this parameter is the most poorly determined. This is one of the main reasons for the extra value of ultra-deep surveys following the OTELO approach.

On the other hand, OTELO covers a relatively small volume of the Universe at $z=0.9$ when compared with previous surveys \citep[see][]{Villar2008,Sobral2015}. For this reason, the probability of detecting luminous sources is small and most of the H$\beta$ emitters detected correspond to low H$\beta$ luminosities, which translates into sub-$L^*$ galaxies, most of them being dwarf systems. The low H$\beta$ luminosity could also correspond to star-forming processes in the final phase, but the short duration of this step implies a low probability of being detected in such a specific evolutive stage.

\subsection{SFR properties}
\label{subsec:SFR}

 \begin{figure*}
   \includegraphics[width=0.5\textwidth]{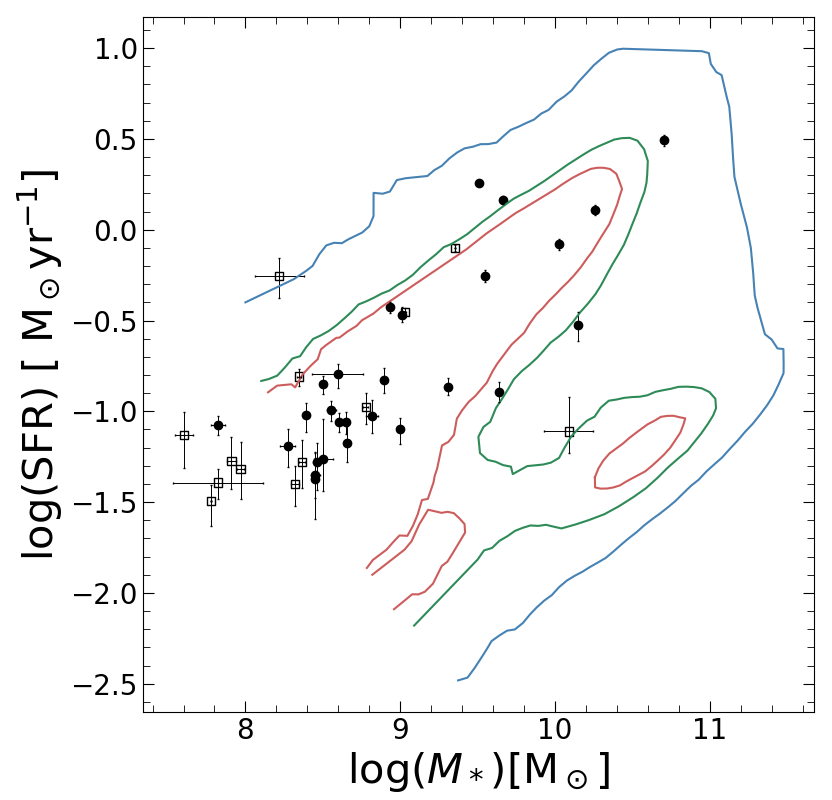}
    \includegraphics[width=0.5\textwidth]{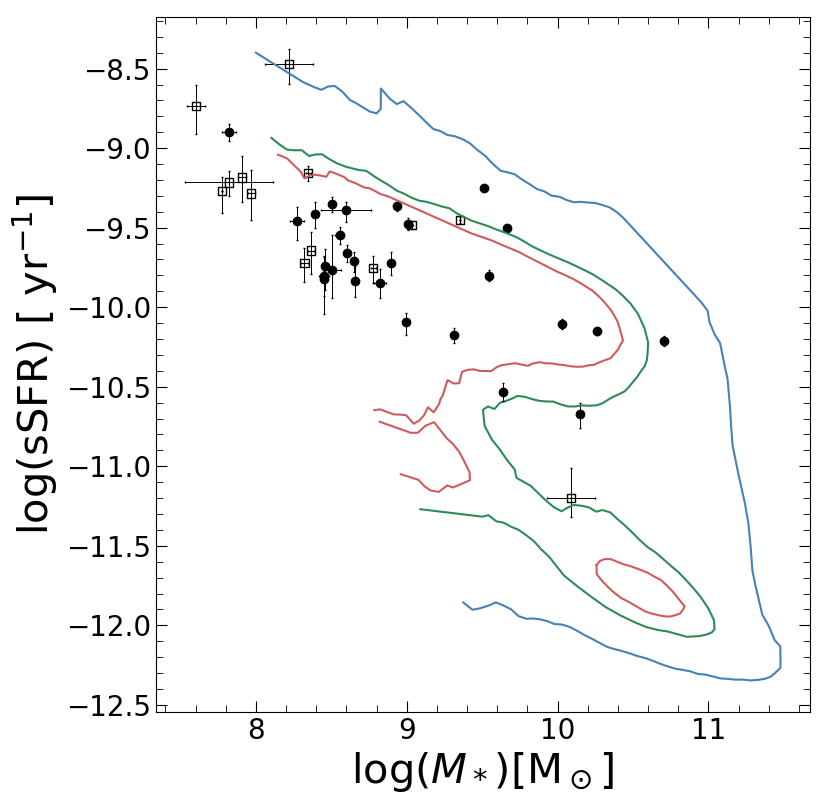}\\
    \includegraphics[width=0.5\textwidth]{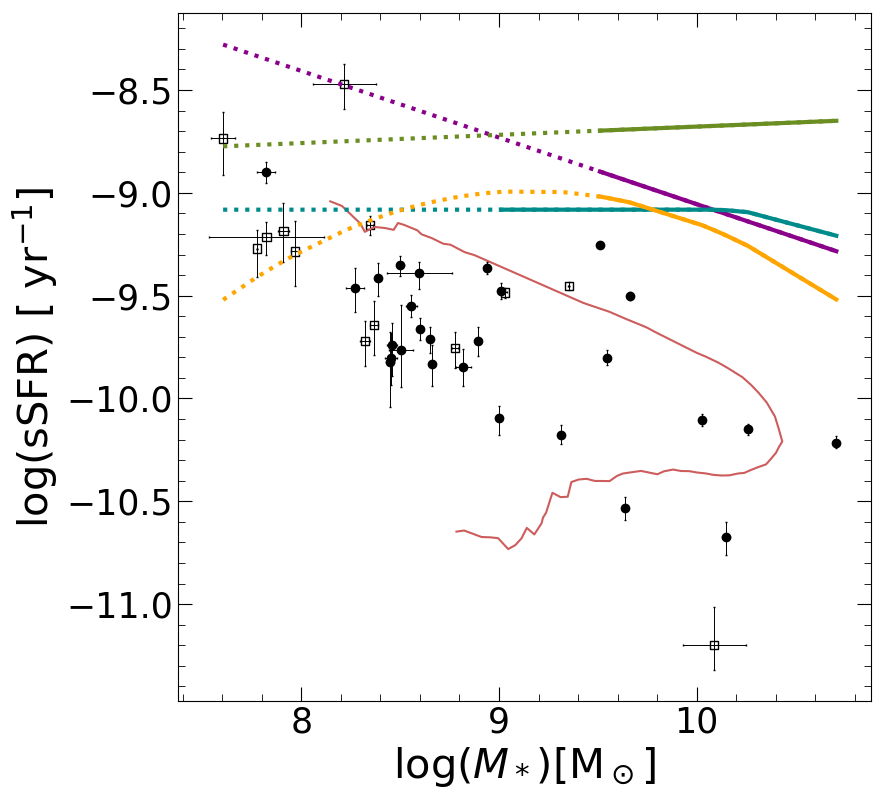}
    \includegraphics[width=0.4\textwidth]{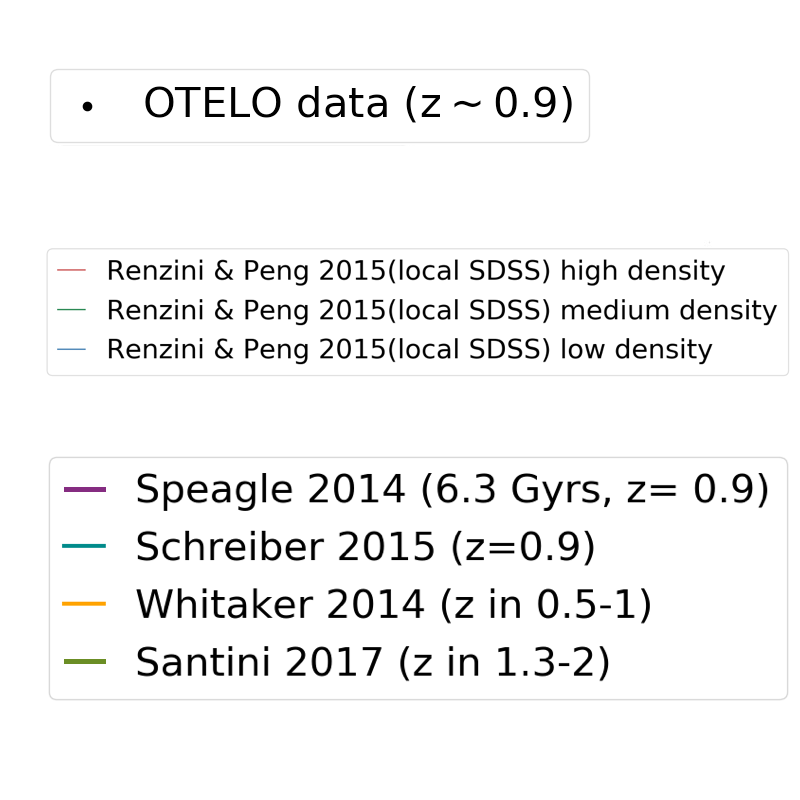}
    \caption{Stellar masses as a function of SFR  (top-left) and sSFR  (top-right) for the \hb sample. Disc and spheroid galaxies are indicated with empty yellow squares and empty grey circles, respectively. Contours correspond to the number density of galaxies from the SDSS database and obtained by \cite{RP2015} at values of $1.2\times 10^5$ (red), $7.0\times 10^4$ (green), and $2.0 \times 10^4$ (blue), clearly showing the position of the SF main sequence for local galaxies. 
    We differentiate the robust sub-sample of 27 galaxies from the remaining 13 with black circles and empty black squares, respectively.  The bottom-left plot clearly shows the local SF main sequence location (the red contour), where most of the galaxies in our sample are located. The plot also shows the different position of the SF main sequence given by different authors. For each MS, we differentiate the mass range used by each author from the extrapolated one by us by plotting the latter with a discontinuous line.}
    \label{fig:sfr}
\end{figure*}

 
 
At $z \sim 0.9$, conventional surveys mainly trace the overall population of disc galaxies in an enhanced star-formation phase that is reflected in cosmic  star formation history of the Universe (see, e.g. \citet{Villar2011}, \citet{Madau2014}). Most of these $L^*$ systems are easy to detect in the NIR, with a  fraction of the population at $z \sim 0.9$ qualifying as luminous infrared galaxies (LIRGs, with stellar masses approximately of $10^{11}\,\mathrm{M}_\odot$) and UV-bright systems. On the contrary, OTELO is tracing a population more similar to late-type and dwarf star-forming galaxies in the Local Universe (i.e. the Magellanic Clouds). This is shown in Figs. \ref{fig:histograms} and \ref{fig:contour} where the peak of observed galaxies is around $\log M_*[\mathrm{M}_\odot] \sim 8.5$. The less massive the galaxies, the larger the uncertainties are in 
mass. Among those objects that could be assigned a morphological type, most were classified as disc and spiral, and they present masses and sSFRs appear in the medium-high range. 

Figure \ref{fig:sfr} shows the stellar mass as a function of the SFR and sSFR in the \hb OTELO sample compared with the local number density distribution of SF galaxies in the SDSS database with $0.02 < z < 0.085$, as obtained by \cite{RP2015}. This figure clearly shows that the \hb OTELO sample at $z\sim 0.9$ (i.e. a universe age of 6.3 Gyr) shares its position with the SF main sequence (hereafter SF-MS) for local galaxies with the exception of one more massive $\log M_* > 10$ located in the green valley (the one with $\log \mathrm{sSFR} \sim 11.2$). It suggests that there is no redshift evolution of the SF-MS in the low--mass regime, in contrast with the observed evolution of the SF-MS for masses with $\log M_* > 9$ \cite[e.g.][ and references therein]{Popess2019}. As a comparison, in bottom left panel of Fig. \ref{fig:sfr}, we plotted SF-MS at $z$ around 0.8 as defined by several authors extrapolated to our observed mass range.\footnote{In \citet{Speagle2014, Schreiber2015} and \citet{Santini2015}, the samples are composed of galaxies with masses larger than $10^{9.5}$ M$\sun$. The results of \citet{Whitaker2014} are based on galaxies with masses larger than $10^9$M$\sun$, but assuming that the correction in stellar masses is only right for values larger than $10^{10}$M$\sun$.
Our galaxies are below the extrapolated position of the low mass SF-MS at $z \sim 0.9$ even taking into account an intrinsic scatter of the SF-MS of about 0.3 dex \cite[e.g.][]{Kurczynski2016}. As an additional test, we computed the sSFR increasing the extinction by a factor of 3 (i.e. to mimic a case where $E(B-V)_\mathrm{Balmer} = 3 \times E(B-V)_\mathrm{SED}$). In this case, 12 galaxies are above the extrapolation of \cite{Schreiber2015} SF-MS, and 28 are below. Actually, a decrease in the  \cite{Schreiber2015} SF-MS of 0.2 dex would be required for it to be at the median value of the sample (included the extra extinction correction).}
In the general scenario of observed downsizing, massive galaxies form most of their stars earlier and on shorter timescales, while less massive galaxies evolve on longer timescales \citep{Cowie1996}. Low-mass star-forming galaxies at $z=0.9$ detected by OTELO present similar properties as low-mass star-forming galaxies in the Local Universe, suggesting that the low mass population of star-forming galaxies is present all along the Universe epochs, with no signs of a favorite epoch of formation or star formation enhancement from $z=1$ to now. It is worth mentioning that most of classic surveys for star-forming galaxies do not properly trace low luminosity star-forming systems. This is reflected in the luminosity function.

\subsection{Luminosity function}
\begin{figure}
\centering
\subfigure[]{\includegraphics[width=0.5\textwidth]{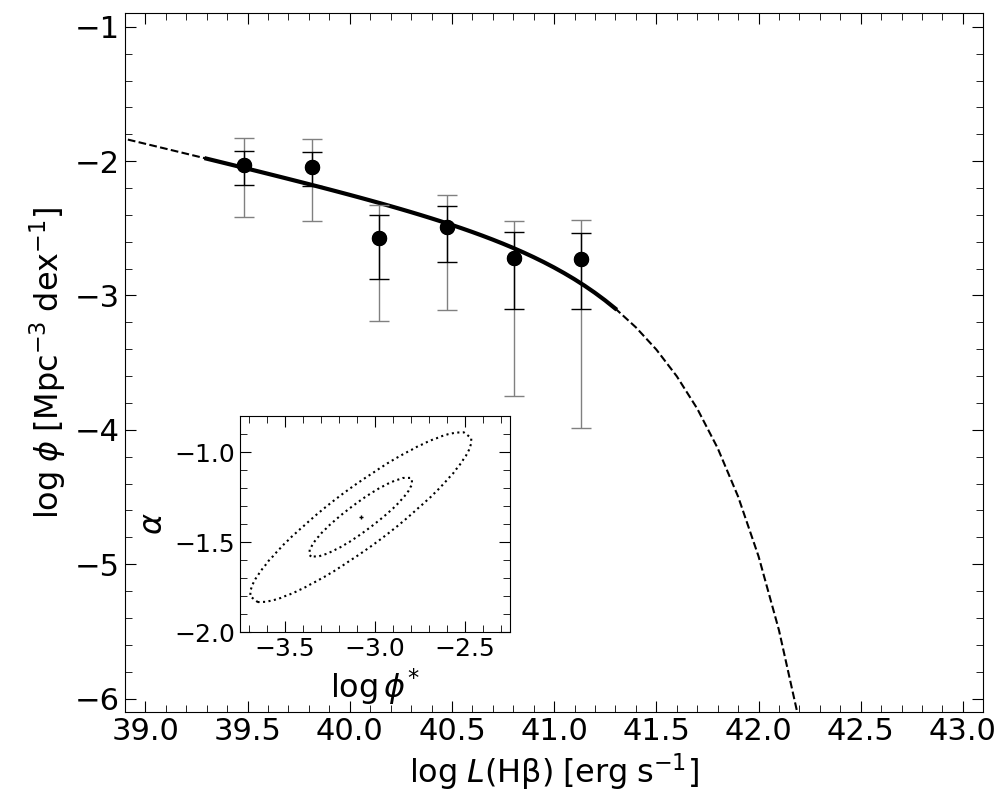}}
\subfigure[]{\includegraphics[width=0.5\textwidth]{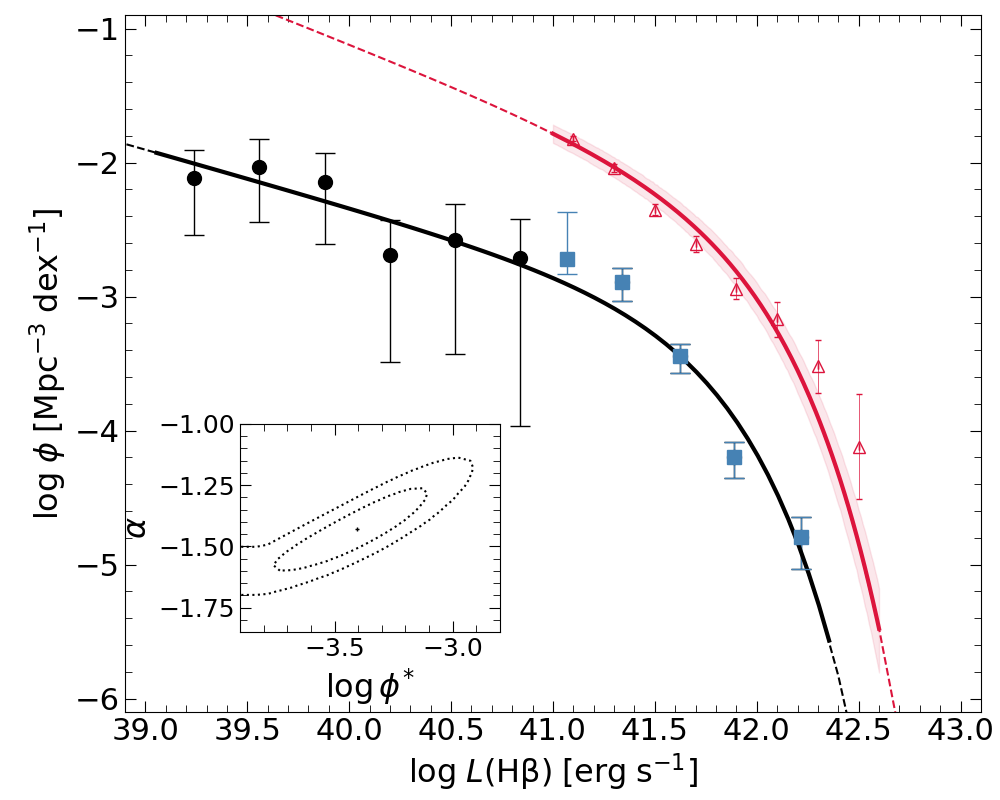}}
\caption{Top figure: Completeness and dust extinction-corrected \hb LF at $z\sim 0.9$. The black dots represent the \hb sample and the black line shows the fitting of this sample. Shorter error bars represent the Poissonian error and the larger error bars have the rest of the uncertainties evenly added in quadrature (see text for details). Bottom figure: OTELO \hb sample (black dots) complemented with high-luminosity data from \citet{Comparat2016} \hb sample at z$\sim$0.8 (blue squares) fitted LF. Red triangles and the red line represent the literature data from the \hb+\oiii{} sample by \citet{Khostovan2015} at $z\sim 0.8$ and its LF fitting, respectively.}
\label{fig:lf}
\end{figure}

The OTELO \hb\ LF best-fit values are shown in Table \ref{funlumtable} and Figure \ref{fig:lf}. As shown there, we performed two different \hb LF fittings.

The first fit was constructed using the dust-corrected $L(\mathrm{H}\beta)$ values. In this case, the $\log L^*$ parameter was fixed to a constant value of 41.34 since our data cover the fainter end of the LF and hence there are not enough to constraint the high and low regimes boundary. This value was drawn from looking at past work from previous narrow-band studies, specifically as a mean value from those used by \citet{Comparat2016} and \citet{Khostovan2015} (see Table \ref{funlumtable}). The values obtained for the log$\phi^*$ and $\alpha$ parameters are -3.08 and -1.36, respectively. As we have explained above, this fitting takes advantage of dust-corrected luminosity values so a comparison with previous works would not be interesting since those used non-corrected data. However, the value obtained for the slope of the faint end, $\alpha$, is very significant since our sample extends further than any other previous work and it would constitute a solid and unpredecented $\alpha$ for dust-corrected \hb LF at $z \sim$0.9. This fitting is shown in the first panel of Figure \ref{fig:lf}.  In Table \ref{funlumtable} we have also included the parameters of the OTELO \oiiis{}{} LF from \cite{Bongiovanni2020} at a similar redshift range as our own.

In an attempt to execute a LF fitting where every parameter  was set free and, at the same time, to extend our sample to the brighter end, we performed a second \hb LF fitting joining the OTELO data with \hb at $z \sim$0.8 data from \citet{Comparat2016}. In this case, we used the non dust-corrected L(\hb) values, since those from \citet{Comparat2016} were not corrected from dust extinction as well. The second panel of Figure \ref{fig:lf} portrays this LF fitting as well as the fit from Khostovan. The sample from \citet{Khostovan2015} contains data from \hb + \oiii{} ELS at $z \sim 0.8$.

The largest relative uncertainties on the Schechter parameters obtained after the non-linear fitting of both LFs correspond to the 68.27\% confidence interval of each parameter. The inset in Figure 5 shows the strong correlation between $\phi*$ and  $\alpha$ parameters.

As stated above, Schechter-LFs parameters are more or less correlated, hence the difficulty when trying to compare different fittings. However, a general agreement is observed between the LF estimates from previous studies and ours. Figure \ref{fig:lf} seemingly shows a significant difference among these different works along the whole LF. In particular, the OTELO LF(\hb) prediction around the surroundings of $L^*$ is about $\sim$1 dex smaller than the model of \citet{Khostovan2015}. Similarly, all of the fitted $\phi^*$ values from the literature are larger than those from our sample. This is likely because the sample of \citet{Khostovan2015} gathers both \oiiis\ and \hb\ emission lines and, as they predict in their study, the bright end is usually dominated by \oiiis\  emitters.
In \citet{Bongiovanni2020}, we obtain the LF of \oiiis\ at $z \sim 0.8$, getting significantly different values for $\phi^*$ and $\alpha$ (see Table \ref{funlumtable}). This implies that the contribution of the oxygen lines to the LF adds different physical parameters to those influencing the H$\beta$ emission. The study presented in this paper shows the LF(H$\beta$) based on the fluxes of this line only, without contamination from \oiiis{}{}.

At low luminosities, the LF obtained from the OTELO \hb\ sample extends about 100 times fainter than the most sensitive extreme observed to date. Hence we are observing faint galaxies that other surveys do not detect. Regarding our $\alpha$ value, it slightly differs from the one used by \citet{Khostovan2015} and by \citet{Comparat2016}, which are $\sim-1.66$ and $\sim-1.51$, respectively. It must be noted, nonetheless, that the $\alpha$ adopted in \citet{Khostovan2015} is a fixed value derived from previous works that do not reach luminosities as low as the \hb OTELO sample does. Hence our results provide a better approach to this estimation. 
Other studies using H$\alpha$ at $z=0.8$ \citep[][]{Sobral2013} reach only $\log L\sim$41.5, and the values of $\alpha$ reported in those works are based in the bright end of the luminosity function. Here we report the LF estimate based in sources up to $\log L\sim$39.5.


Moreover, \citet{Drake2013} infer that the detection fraction of ELSs strongly determines the faint-end slope of the LF, but also that the $\alpha$ value is sensitive to the adopted limit of EW of a typical NB survey. Accordingly, since the EW lower limit of OTELO data is around 6~\AA, we can conclude the faint-end slope value provided by our \hb\ fit is robust enough.

\section{Summary and conclusions}
OTELO is a 2D spectroscopic blind survey, with a spectral resolution of R=700, covering a field in the EGS of $7.5\times 7.4$ arcmin$^2$ area. Using the OSIRIS TF, a window of 230 \AA\ centred at 9175 \AA\ was scanned with 36 
slices evenly spaced by 6 \AA. OTELO obtained photometric data at consecutive and overlapping wavelength ranges (pseudo-spectra) of all ELS in the field, hence covering a wide range of volumes between  $z=0.4$ and 6. The final product is a set of astrometry-corrected and flux-calibrated images of each slice as well as a pseudo-spectrum for every source of the field, obtained by doing aperture photometry in the images. Details on the survey strategy, data reduction, and main products are provided in OTELO-I.

In this paper, we have exploited the scientific potential of the selection of ELSs detailed in OTELO-I and focused on \hb\ emitters. The selection procedure for this very sample provided 87 preliminary ELSs in the redshift window around $z=0.88$. From this selection, 41 objects constitute the final sample. We performed a deconvolution of their emission lines in order to obtain accurate redshifts, line fluxes, and observed EW. The \hb sample is distributed in a redshift range between 0.86 and 0.9, with a limiting line flux of $\sim 1\times 10^{-18}$ erg s$^{-1}$ cm$^2$ with an EW as low as $\sim 9$ \AA. Most of the morphologically classified \hb\ ELSs are disc-like galaxies (76\%), and stellar masses range between $10^{7.6}$ - $10^{10.7}\, \mathrm{M}_{\odot}$, with 90\% of the sample in the low-mass galaxy population ($\mathrm{M}_*< 10^{10}\mathrm{M}_\odot$). After searching for AGN host candidates as described in Sect.\ref{sec:AGN}, only one source was classified as such. The OTELO survey hence provides high sensitivity to the detection of faint SFGs and a very significant minimum line flux.

The SFR was derived from dust and stellar absorption-corrected \hb luminosity. The SFR result places our sample in the SFR main sequence according to Santini et al.(2017), and it induces a similarity between this sources and dwarf star-forming galaxies in the Local Universe. Being our sample mostly formed by galaxies with masses below 10$^{9.5}$M$_\sun$ and under the assumption of a moderate correction of  $E(B-V)_\mathrm{Balmer} \leq 3 \times E(B-V)_\mathrm{SED}$, we conclude that our data are compatible with no evolution in the SFR of low-mass galaxies.

We computed the luminosity for each galaxy of our ELS sample from the fluxes obtained from inverse deconvolution. In sampling a co-moving volume of $\sim$5190 Mpc$^3$ and mainly taking the sources of uncertainties into account (primarily CV effects), we obtained the observed non dust-corrected LF of the \hb sample. The sample produced an unprecedented sampling of the faint end of the \hb LF as it is 100 times fainter than the extreme reached by other surveys to date. This dust-corrected OTELO \hb LF delivered the following Schechter parameters: $\log \phi^* = -3.08 \pm 0.19$, $\log L^* = 41.34$, and $\alpha = -1.36 \pm 0.15$. A second LF fitting was made by extending the bright end of our non dust-corrected sample with the \hb+\oiii\ data from \citet{Comparat2016} and the parameters for the best fit for this junction were: $\log \phi^* = -3.40 \pm 0.2$, $\log L^* = 41.65 \pm 0.1$, and $\alpha=-1.43 \pm 0.12$. This faint-end slope value is consistent with previous similar works, and it represents the most robust alpha estimation at $z\sim$0.8 published so far, based on the faintest isolated \hb\ (and \ha) detections to date. In this sense, OTELO is complementary to other surveys, but it provides an unreached sensitivity to smaller, faint galaxies.

\begin{acknowledgements}
This paper is dedicated to the memory of our dear friend and colleague Hector Castañeda,  unfortunately died on 19 Nov 2020. This  work  was  supported  by  the  Spanish  Ministry  of  Economy  and
Competitiveness  (MINECO) under  the  grants  AYA2014\,-\,58861\,-\,C3\,-\,1\,-\,P,
AYA2014\,-\,58861\,-\,C3\,-\,2\,-\,P,  AYA2014\,-\,58861\,-\,C3\,-\,3\,-\,P, 
AYA2013\,-\,46724\,-\,P, AYA2017\,-\,88007\,–\,C3\,–\,1\,–\,P, AYA2017\,–\,88007\,–\,C3\,–\,2, MDM-2017-0737 (Unidad de Excelencia María de Maeztu, CAB).  APG acknowledge support from ESA through the Faculty of the European
Space Astronomy Centre (ESAC) - Funding reference ESAC\_549/2019.
Based on observations made with the Gran Telescopio Canarias (GTC), installed in the
Spanish Observatorio del Roque de los Muchachos of the Instituto de Astrof\'isica de
Canarias, in the island of La Palma.

This study makes use of data from AEGIS, a multiwavelength sky survey conducted with the
Chandra, GALEX, Hubble, Keck, CFHT, MMT, Subaru, Palomar, Spitzer, VLA, and other telescopes
and supported in part by the NSF, NASA, and the STFC.

Based  on  observations  obtained  with  MegaPrime/MegaCam,  a  joint  project  of  CFHT  and
CEA/IRFU, at the Canada-France-Hawaii Telescope (CFHT) which is operated by the National
Research Council (NRC) of Canada, the Institut National des Science de l'Univers of the
Centre National de la Recherche Scientifique (CNRS) of France, and the University of
Hawaii.  This work is based in part on data products produced at Terapix available at
the Canadian Astronomy Data Centre as part of the Canada-France-Hawaii Telescope Legacy
Survey, a collaborative project of NRC and CNRS.

Based on observations obtained with WIRCam, a joint project of CFHT,Taiwan, Korea, Canada,
France, at the Canada-France-Hawaii Telescope (CFHT) which is operated by the National
Research Council (NRC) of Canada, the Institute National des Sciences de l'Univers of the
Centre National de la Recherche Scientifique of France, and the University of Hawaii.
This work is based in part on data products produced at TERAPIX, the WIRDS (WIRcam Deep
Survey) consortium, and the Canadian Astronomy Data Centre. This research was supported by
a grant from the Agence Nationale de la Recherche ANR-07-BLAN-0228

\end{acknowledgements}

\bibliographystyle{aa} 
\bibliography{biblio}

\begin{thebibliography}{61}
\expandafter\ifx\csname natexlab\endcsname\relax\def\natexlab#1{#1}\fi

\bibitem[{{Alvarez} {et~al.}(1998){Alvarez}, {Rodr{\'\i}guez Espinosa}, \&
  {S{\'a}nchez}}]{Alvarez98}
{Alvarez}, P., {Rodr{\'\i}guez Espinosa}, J.~M., \& {S{\'a}nchez}, F. 1998,
  \nar, 42, 553

\bibitem[{{Arnouts} {et~al.}(1999){Arnouts}, {Cristiani}, {Moscardini},
  {Matarrese}, {Lucchin}, {Fontana}, \& {Giallongo}}]{Arnouts1999}
{Arnouts}, S., {Cristiani}, S., {Moscardini}, L., {et~al.} 1999, \mnras, 310,
  540

\bibitem[{{Bluck} {et~al.}(2014){Bluck}, {Mendel}, {Ellison}, {Moreno},
  {Simard}, {Patton}, \& {Starkenburg}}]{Bluck2014}
{Bluck}, A. F.~L., {Mendel}, J.~T., {Ellison}, S.~L., {et~al.} 2014, \mnras,
  441, 599

\bibitem[{{Bongiorno} {et~al.}(2010){Bongiorno}, {Mignoli}, {Zamorani},
  {Lamareille}, {Lanzuisi}, {Miyaji}, {Bolzonella}, {Carollo}, {Contini},
  {Kneib}, {Le F{\`e}vre}, {Lilly}, {Mainieri}, {Renzini}, {Scodeggio},
  {Bardelli}, {Brusa}, {Caputi}, {Civano}, {Coppa}, {Cucciati}, {de la Torre},
  {de Ravel}, {Franzetti}, {Garilli}, {Halliday}, {Hasinger}, {Koekemoer},
  {Iovino}, {Kampczyk}, {Knobel}, {Kova{\v{c}}}, {Le Borgne}, {Le Brun},
  {Maier}, {Merloni}, {Nair}, {Pello}, {Peng}, {Perez Montero}, {Ricciardelli},
  {Salvato}, {Silverman}, {Tanaka}, {Tasca}, {Tresse}, {Vergani}, {Zucca},
  {Abbas}, {Bottini}, {Cappi}, {Cassata}, {Cimatti}, {Guzzo}, {Leauthaud},
  {Maccagni}, {Marinoni}, {McCracken}, {Memeo}, {Meneux}, {Oesch}, {Porciani},
  {Pozzetti}, \& {Scaramella}}]{Bongiorno2010}
{Bongiorno}, A., {Mignoli}, M., {Zamorani}, G., {et~al.} 2010, \aap, 510, A56

\bibitem[{{Bongiovanni} {et~al.}(2019){Bongiovanni}, {Ram{\'o}n-P{\'e}rez},
  {P{\'e}rez Garc{\'\i}a}, {Cepa}, {Cervi{\~n}o}, {Nadolny}, {P{\'e}rez
  Mart{\'\i}nez}, {Alfaro}, {Casta{\~n}eda}, {de Diego}, {Ederoclite},
  {Fern{\'a}ndez-Lorenzo}, {Gallego}, {Gonz{\'a}lez}, {Gonz{\'a}lez-Serrano},
  {Lara-L{\'o}pez}, {Oteo G{\'o}mez}, {Padilla Torres}, {Pintos-Castro},
  {Povi{\'c}}, {S{\'a}nchez-Portal}, {Jones}, {Bland-Hawthorn}, \&
  {Cabrera-Lavers}}]{Bongiovanni2019}
{Bongiovanni}, {\'A}., {Ram{\'o}n-P{\'e}rez}, M., {P{\'e}rez Garc{\'\i}a},
  A.~M., {et~al.} 2019, \aap, 631, A9

\bibitem[{{Bongiovanni} {et~al.}(2020){Bongiovanni}, {Ram{\'o}n-P{\'e}rez},
  {P{\'e}rez Garc{\'\i}a}, {Cervi{\~n}o}, {Cepa}, {Nadolny}, {P{\'e}rez
  Mart{\'\i}nez}, {Alfaro}, {Casta{\~n}eda}, {Cedr{\'e}s}, {de Diego},
  {Ederoclite}, {Fern{\'a}ndez-Lorenzo}, {Gallego}, {de Jes{\'u}s
  Gonz{\'a}lez}, {Gonz{\'a}lez-Serrano}, {Lara-L{\'o}pez}, {Oteo G{\'o}mez},
  {Padilla Torres}, {Pintos-Castro}, {Povi{\'c}}, {S{\'a}nchez-Portal}, {Heath
  Jones}, {Bland-Hawthorn}, \& {Cabrera-Lavers}}]{Bongiovanni2020}
{Bongiovanni}, {\'A}., {Ram{\'o}n-P{\'e}rez}, M., {P{\'e}rez Garc{\'\i}a},
  A.~M., {et~al.} 2020, \aap, 635, A35

\bibitem[{{Bouch{\'e}} {et~al.}(2010){Bouch{\'e}}, {Dekel}, {Genzel}, {Genel},
  {Cresci}, {F{\"o}rster Schreiber}, {Shapiro}, {Davies}, \&
  {Tacconi}}]{Bouche10}
{Bouch{\'e}}, N., {Dekel}, A., {Genzel}, R., {et~al.} 2010, \apj, 718, 1001

\bibitem[{{Cardelli} {et~al.}(1989){Cardelli}, {Clayton}, \&
  {Mathis}}]{Cardelli1989}
{Cardelli}, J.~A., {Clayton}, G.~C., \& {Mathis}, J.~S. 1989, \apj, 345, 245

\bibitem[{{Cepa} {et~al.}(2003){Cepa}, {Alfaro}, {Bland-Hawthorn},
  {Casta{\~n}eda}, {Ga-Llego}, {Gonz{\'a}lez-Serrano}, {Gonz{\'a}lez}, \&
  {S{\'a}nchez-Portal}}]{Cepa98}
{Cepa}, J., {Alfaro}, E.~J., {Bland-Hawthorn}, J., {et~al.} 2003, in Revista
  Mexicana de Astronomia y Astrofisica Conference Series, Vol.~16, Revista
  Mexicana de Astronomia y Astrofisica Conference Series, ed. J.~M. {Rodriguez
  Espinoza}, F.~{Garzon Lopez}, \& V.~{Melo Martin}, 64--68

\bibitem[{{Cervi{\~n}o} {et~al.}(2016){Cervi{\~n}o}, {Bongiovanni}, \&
  {Hidalgo}}]{Cervino2016}
{Cervi{\~n}o}, M., {Bongiovanni}, A., \& {Hidalgo}, S. 2016, \aap, 589, A108

\bibitem[{{Chiang} {et~al.}(2019){Chiang}, {Goto}, {Hashimoto}, {Kim},
  {Matsuhara}, \& {Oi}}]{Chiang2019}
{Chiang}, C.-Y., {Goto}, T., {Hashimoto}, T., {et~al.} 2019, \pasj, 71, 31

\bibitem[{{Coleman} {et~al.}(1980){Coleman}, {Wu}, \& {Weedman}}]{Coleman1980}
{Coleman}, G.~D., {Wu}, C.~C., \& {Weedman}, D.~W. 1980, \apjs, 43, 393

\bibitem[{{Comparat} {et~al.}(2016){Comparat}, {Zhu}, {Gonzalez-Perez},
  {Norberg}, {Newman}, {Tresse}, {Richard}, {Yepes}, {Kneib}, {Raichoor},
  {Prada}, {Maraston}, {Y{\`e}che}, {Delubac}, \& {Jullo}}]{Comparat2016}
{Comparat}, J., {Zhu}, G., {Gonzalez-Perez}, V., {et~al.} 2016, \mnras, 461,
  1076

\bibitem[{{Coughlin} {et~al.}(2018){Coughlin}, {Rhoads}, {Malhotra}, {Probst},
  {Swaters}, {Tilvi}, {Zheng}, {Finkelstein}, {Hibon}, {Mobasher}, {Jiang},
  {Joshi}, {Pharo}, {Veilleux}, {Wang}, {Yang}, \& {Zabl}}]{Coughlin2018}
{Coughlin}, A., {Rhoads}, J.~E., {Malhotra}, S., {et~al.} 2018, \apj, 858, 96

\bibitem[{{Cowie} {et~al.}(1996){Cowie}, {Songaila}, {Hu}, \&
  {Cohen}}]{Cowie1996}
{Cowie}, L.~L., {Songaila}, A., {Hu}, E.~M., \& {Cohen}, J.~G. 1996, \aj, 112,
  839

\bibitem[{{Donley} {et~al.}(2012){Donley}, {Koekemoer}, {Brusa}, {Capak},
  {Cardamone}, {Civano}, {Ilbert}, {Impey}, {Kartaltepe}, {Miyaji}, {Salvato},
  {Sanders}, {Trump}, \& {Zamorani}}]{Donley2012}
{Donley}, J.~L., {Koekemoer}, A.~M., {Brusa}, M., {et~al.} 2012, \apj, 748, 142

\bibitem[{{Drake} {et~al.}(2013){Drake}, {Simpson}, {Collins}, {James},
  {Baldry}, {Ouchi}, {Jarvis}, {Bonfield}, {Ono}, {Best}, {Dalton}, {Dunlop},
  {McLure}, \& {Smith}}]{Drake2013}
{Drake}, A.~B., {Simpson}, C., {Collins}, C.~A., {et~al.} 2013, \mnras, 433,
  796

\bibitem[{{Driver} {et~al.}(2009){Driver}, {Norberg}, {Baldry}, {Bamford},
  {Hopkins}, {Liske}, {Loveday}, {Peacock}, {Hill}, {Kelvin}, {Robotham},
  {Cross}, {Parkinson}, {Prescott}, {Conselice}, {Dunne}, {Brough}, {Jones},
  {Sharp}, {van Kampen}, {Oliver}, {Roseboom}, {Bland -Hawthorn}, {Croom},
  {Ellis}, {Cameron}, {Cole}, {Frenk}, {Couch}, {Graham}, {Proctor}, {De
  Propris}, {Doyle}, {Edmondson}, {Nichol}, {Thomas}, {Eales}, {Jarvis},
  {Kuijken}, {Lahav}, {Madore}, {Seibert}, {Meyer}, {Staveley-Smith},
  {Phillipps}, {Popescu}, {Sansom}, {Sutherland}, {Tuffs}, \&
  {Warren}}]{Driver2009}
{Driver}, S.~P., {Norberg}, P., {Baldry}, I.~K., {et~al.} 2009, Astronomy and
  Geophysics, 50, 5.12

\bibitem[{{Elbaz} {et~al.}(2007){Elbaz}, {Daddi}, {Le Borgne}, {Dickinson},
  {Alexander}, {Chary}, {Starck}, {Brand t}, {Kitzbichler}, {MacDonald},
  {Nonino}, {Popesso}, {Stern}, \& {Vanzella}}]{Elbaz07}
{Elbaz}, D., {Daddi}, E., {Le Borgne}, D., {et~al.} 2007, \aap, 468, 33

\bibitem[{{Elmegreen} {et~al.}(2007){Elmegreen}, {Elmegreen}, {Ravindranath},
  \& {Coe}}]{Elmegreen2007a}
{Elmegreen}, D.~M., {Elmegreen}, B.~G., {Ravindranath}, S., \& {Coe}, D.~A.
  2007, \apj, 658, 763

\bibitem[{{Fujita} {et~al.}(2003){Fujita}, {Ajiki}, {Shioya}, {Nagao},
  {Murayama}, {Taniguchi}, {Umeda}, {Yamada}, {Yagi}, {Okamura}, \&
  {Komiyama}}]{Fujita2003}
{Fujita}, S.~S., {Ajiki}, M., {Shioya}, Y., {et~al.} 2003, \apjl, 586, L115

\bibitem[{{Gallego} {et~al.}(1995){Gallego}, {Zamorano}, {Aragon-Salamanca}, \&
  {Rego}}]{Gallego1995}
{Gallego}, J., {Zamorano}, J., {Aragon-Salamanca}, A., \& {Rego}, M. 1995,
  \apjl, 455, L1

\bibitem[{{Geach} {et~al.}(2008){Geach}, {Smail}, {Best}, {Kurk}, {Casali},
  {Ivison}, \& {Coppin}}]{Geach2008}
{Geach}, J.~E., {Smail}, I., {Best}, P.~N., {et~al.} 2008, \mnras, 388, 1473

\bibitem[{{Hayashi} {et~al.}(2018){Hayashi}, {Tanaka}, {Shimakawa}, {Furusawa},
  {Momose}, {Koyama}, {Silverman}, {Kodama}, {Komiyama}, {Leauthaud}, {Lin},
  {Miyazaki}, {Nagao}, {Nishizawa}, {Ouchi}, {Shibuya}, {Tadaki}, \&
  {Yabe}}]{Hayashi2018}
{Hayashi}, M., {Tanaka}, M., {Shimakawa}, R., {et~al.} 2018, \pasj, 70, S17

\bibitem[{{Hopkins} {et~al.}(2013){Hopkins}, {Driver}, {Brough}, {Owers},
  {Bauer}, {Gunawardhana}, {Cluver}, {Colless}, {Foster}, {Lara-L{\'o}pez},
  {Roseboom}, {Sharp}, {Steele}, {Thomas}, {Baldry}, {Brown}, {Liske},
  {Norberg}, {Robotham}, {Bamford}, {Bland-Hawthorn}, {Drinkwater}, {Loveday},
  {Meyer}, {Peacock}, {Tuffs}, {Agius}, {Alpaslan}, {Andrae}, {Cameron},
  {Cole}, {Ching}, {Christodoulou}, {Conselice}, {Croom}, {Cross}, {De
  Propris}, {Delhaize}, {Dunne}, {Eales}, {Ellis}, {Frenk}, {Graham},
  {Grootes}, {H{\"a}u{\ss}ler}, {Heymans}, {Hill}, {Hoyle}, {Hudson}, {Jarvis},
  {Johansson}, {Jones}, {van Kampen}, {Kelvin}, {Kuijken},
  {L{\'o}pez-S{\'a}nchez}, {Maddox}, {Madore}, {Maraston}, {McNaught-Roberts},
  {Nichol}, {Oliver}, {Parkinson}, {Penny}, {Phillipps}, {Pimbblet}, {Ponman},
  {Popescu}, {Prescott}, {Proctor}, {Sadler}, {Sansom}, {Seibert},
  {Staveley-Smith}, {Sutherland }, {Taylor}, {Van Waerbeke},
  {V{\'a}zquez-Mata}, {Warren}, {Wijesinghe}, {Wild}, \&
  {Wilkins}}]{Hopkins2013}
{Hopkins}, A.~M., {Driver}, S.~P., {Brough}, S., {et~al.} 2013, \mnras, 430,
  2047

\bibitem[{{Hopkins} {et~al.}(2003){Hopkins}, {Miller}, {Nichol}, {Connolly},
  {Bernardi}, {G{\'o}mez}, {Goto}, {Tremonti}, {Brinkmann}, {Ivezi{\'c}}, \&
  {Lamb}}]{Hopkins2003}
{Hopkins}, A.~M., {Miller}, C.~J., {Nichol}, R.~C., {et~al.} 2003, \apj, 599,
  971

\bibitem[{{Ilbert} {et~al.}(2006){Ilbert}, {Arnouts}, {McCracken},
  {Bolzonella}, {Bertin}, {Le F{\`e}vre}, {Mellier}, {Zamorani}, {Pell{\`o}},
  {Iovino}, {Tresse}, {Le Brun}, {Bottini}, {Garilli}, {Maccagni}, {Picat},
  {Scaramella}, {Scodeggio}, {Vettolani}, {Zanichelli}, {Adami}, {Bardelli},
  {Cappi}, {Charlot}, {Ciliegi}, {Contini}, {Cucciati}, {Foucaud}, {Franzetti},
  {Gavignaud}, {Guzzo}, {Marano}, {Marinoni}, {Mazure}, {Meneux}, {Merighi},
  {Paltani}, {Pollo}, {Pozzetti}, {Radovich}, {Zucca}, {Bondi}, {Bongiorno},
  {Busarello}, {de La Torre}, {Gregorini}, {Lamareille}, {Mathez}, {Merluzzi},
  {Ripepi}, {Rizzo}, \& {Vergani}}]{Ilbert2006}
{Ilbert}, O., {Arnouts}, S., {McCracken}, H.~J., {et~al.} 2006, \aap, 457, 841

\bibitem[{{Kartaltepe} {et~al.}(2015){Kartaltepe}, {Mozena}, {Kocevski},
  {McIntosh}, {Lotz}, {Bell}, {Faber}, {Ferguson}, {Koo}, {Bassett}, {Bernyk},
  {Blancato}, {Bournaud}, {Cassata}, {Castellano}, {Cheung}, {Conselice},
  {Croton}, {Dahlen}, {de Mello}, {DeGroot}, {Donley}, {Guedes}, {Grogin},
  {Hathi}, {Hilton}, {Hollon}, {Koekemoer}, {Liu}, {Lucas}, {Martig},
  {McGrath}, {McPartland}, {Mobasher}, {Morlock}, {O'Leary}, {Peth}, {Pforr},
  {Pillepich}, {Rosario}, {Soto}, {Straughn}, {Telford}, {Sunnquist}, {Trump},
  {Weiner}, {Wuyts}, {Inami}, {Kassin}, {Lani}, {Poole}, \&
  {Rizer}}]{Kartaltepe2015}
{Kartaltepe}, J.~S., {Mozena}, M., {Kocevski}, D., {et~al.} 2015, \apjs, 221,
  11

\bibitem[{{Kennicutt}(1998)}]{Kennicutt1998}
{Kennicutt}, Robert~C., J. 1998, \araa, 36, 189

\bibitem[{{Kennicutt} \& {Evans}(2012)}]{Kennicutt2012}
{Kennicutt}, R.~C. \& {Evans}, N.~J. 2012, \araa, 50, 531

\bibitem[{{Khostovan} {et~al.}(2015){Khostovan}, {Sobral}, {Mobasher}, {Best},
  {Smail}, {Stott}, {Hemmati}, \& {Nayyeri}}]{Khostovan2015}
{Khostovan}, A.~A., {Sobral}, D., {Mobasher}, B., {et~al.} 2015, \mnras, 452,
  3948

\bibitem[{{Kinney} {et~al.}(1996){Kinney}, {Calzetti}, {Bohlin}, {McQuade},
  {Storchi-Bergmann}, \& {Schmitt}}]{Kinney1996}
{Kinney}, A.~L., {Calzetti}, D., {Bohlin}, R.~C., {et~al.} 1996, \apj, 467, 38

\bibitem[{{Kroupa}(2001)}]{Kroupa2001}
{Kroupa}, P. 2001, \mnras, 322, 231

\bibitem[{{Kurczynski} {et~al.}(2016){Kurczynski}, {Gawiser}, {Acquaviva},
  {Bell}, {Dekel}, {de Mello}, {Ferguson}, {Gardner}, {Grogin}, {Guo},
  {Hopkins}, {Koekemoer}, {Koo}, {Lee}, {Mobasher}, {Primack}, {Rafelski},
  {Soto}, \& {Teplitz}}]{Kurczynski2016}
{Kurczynski}, P., {Gawiser}, E., {Acquaviva}, V., {et~al.} 2016, \apjl, 820, L1

\bibitem[{{Le F{\`e}vre} {et~al.}(2013){Le F{\`e}vre}, {Cassata}, {Cucciati},
  {Garilli}, {Ilbert}, {Le Brun}, {Maccagni}, {Moreau}, {Scodeggio}, {Tresse},
  {Zamorani}, {Adami}, {Arnouts}, {Bardelli}, {Bolzonella}, {Bondi},
  {Bongiorno}, {Bottini}, {Cappi}, {Charlot}, {Ciliegi}, {Contini}, {de la
  Torre}, {Foucaud}, {Franzetti}, {Gavignaud}, {Guzzo}, {Iovino}, {Lemaux},
  {L{\'o}pez-Sanjuan}, {McCracken}, {Marano}, {Marinoni}, {Mazure}, {Mellier},
  {Merighi}, {Merluzzi}, {Paltani}, {Pell{\`o}}, {Pollo}, {Pozzetti},
  {Scaramella}, {Tasca}, {Vergani}, {Vettolani}, {Zanichelli}, \&
  {Zucca}}]{LeFevre2013}
{Le F{\`e}vre}, O., {Cassata}, P., {Cucciati}, O., {et~al.} 2013, \aap, 559,
  A14

\bibitem[{{L{\'o}pez-Sanjuan} {et~al.}(2019){L{\'o}pez-Sanjuan},
  {D{\'\i}az-Garc{\'\i}a}, {Cenarro}, {Fern{\'a}ndez-Soto}, {Viironen},
  {Molino}, {Ben{\'\i}tez}, {Crist{\'o}bal-Hornillos}, {Moles}, {Varela},
  {Arnalte-Mur}, {Ascaso}, {Castander}, {Cervi{\~n}o}, {Gonz{\'a}lez Delgado},
  {Husillos}, {M{\'a}rquez}, {Masegosa}, {Del Olmo}, {Povi{\'c}}, \&
  {Perea}}]{LopezSanJuan2019}
{L{\'o}pez-Sanjuan}, C., {D{\'\i}az-Garc{\'\i}a}, L.~A., {Cenarro}, A.~J.,
  {et~al.} 2019, \aap, 622, A51

\bibitem[{{Madau} \& {Dickinson}(2014)}]{Madau2014}
{Madau}, P. \& {Dickinson}, M. 2014, \araa, 52, 415

\bibitem[{{Matthee} {et~al.}(2017){Matthee}, {Sobral}, {Best}, {Smail}, {Bian},
  {Darvish}, {R{\"o}ttgering}, \& {Fan}}]{Matthee2017}
{Matthee}, J., {Sobral}, D., {Best}, P., {et~al.} 2017, \mnras, 471, 629

\bibitem[{{Moster} {et~al.}(2011){Moster}, {Somerville}, {Newman}, \&
  {Rix}}]{Moster2011}
{Moster}, B.~P., {Somerville}, R.~S., {Newman}, J.~A., \& {Rix}, H.-W. 2011,
  \apj, 731, 113

\bibitem[{{Nadolny} {et~al.}(2021){Nadolny}, {Bongiovanni}, {Cepa},
  {Cervi{\~n}o}, {P{\'e}rez Garc{\'\i}a}, {Povi{\'c}}, {P{\'e}rez
  Mart{\'\i}nez}, {S{\'a}nchez-Portal}, {de Diego}, {Pintos-Castro}, {Alfaro},
  {Casta{\~n}eda}, {Gallego}, {Jes{\'u}s Gonz{\'a}lez}, {Ignacio
  Gonz{\'a}lez-Serrano}, {Lara-L{\'o}pez}, \& {Padilla Torres}}]{Nadolny2021}
{Nadolny}, J., {Bongiovanni}, {\'A}., {Cepa}, J., {et~al.} 2021, \aap, 647, A89

\bibitem[{{Nadolny} {et~al.}(2020){Nadolny}, {Lara-L{\'o}pez}, {Cervi{\~n}o},
  {Bongiovanni}, {Cepa}, {de Diego}, {P{\'e}rez Garc{\'\i}a}, {P{\'e}rez
  Mart{\'\i}nez}, {S{\'a}nchez-Portal}, {Alfaro}, {Casta{\~n}eda}, {Gallego},
  {Gonz{\'a}lez}, {Gonz{\'a}lez-Serrano}, {Padilla Torres}, {Pintos-Castro}, \&
  {Povi{\'c}}}]{Nadolny2020}
{Nadolny}, J., {Lara-L{\'o}pez}, M.~A., {Cervi{\~n}o}, M., {et~al.} 2020, \aap,
  636, A84

\bibitem[{{Newman} {et~al.}(2013){Newman}, {Cooper}, {Davis}, {Faber}, {Coil},
  {Guhathakurta}, {Koo}, {Phillips}, {Conroy}, {Dutton}, {Finkbeiner}, {Gerke},
  {Rosario}, {Weiner}, {Willmer}, {Yan}, {Harker}, {Kassin}, {Konidaris},
  {Lai}, {Madgwick}, {Noeske}, {Wirth}, {Connolly}, {Kaiser}, {Kirby},
  {Lemaux}, {Lin}, {Lotz}, {Luppino}, {Marinoni}, {Matthews}, {Metevier}, \&
  {Schiavon}}]{Newman2013}
{Newman}, J.~A., {Cooper}, M.~C., {Davis}, M., {et~al.} 2013, \apjs, 208, 5

\bibitem[{{Noeske} {et~al.}(2007){Noeske}, {Weiner}, {Faber}, {Papovich},
  {Koo}, {Somerville}, {Bundy}, {Conselice}, {Newman}, {Schiminovich}, {Le
  Floc'h}, {Coil}, {Rieke}, {Lotz}, {Primack}, {Barmby}, {Cooper}, {Davis},
  {Ellis}, {Fazio}, {Guhathakurta}, {Huang}, {Kassin}, {Martin}, {Phillips},
  {Rich}, {Small}, {Willmer}, \& {Wilson}}]{Noeske07}
{Noeske}, K.~G., {Weiner}, B.~J., {Faber}, S.~M., {et~al.} 2007, \apjl, 660,
  L43

\bibitem[{{P\'erez-Mart\'inez}(2016)}]{PerezMartinez2016}
{P\'erez-Mart\'inez}, R.~M. 2016, {PhD "From Xrays to Far Infrared: Galaxy
  Cluster ZwCL0024+1652 under the multiwavelength limelight", Universidad
  Complutense de Madrid}

\bibitem[{{Popesso} {et~al.}(2019){Popesso}, {Morselli}, {Concas}, {Schreiber},
  {Rodighiero}, {Cresci}, {Belli}, {Ilbert}, {Erfanianfar}, {Mancini}, {Inami},
  {Dickinson}, {Pannella}, \& {Elbaz}}]{Popess2019}
{Popesso}, P., {Morselli}, L., {Concas}, A., {et~al.} 2019, \mnras, 490, 5285

\bibitem[{{Ram{\'o}n-P{\'e}rez}
  {et~al.}(2019{\natexlab{a}}){Ram{\'o}n-P{\'e}rez}, {Bongiovanni}, {P{\'e}rez
  Garc{\'\i}a}, {Cepa}, {Lara-L{\'o}pez}, {de Diego}, {Alfaro},
  {Casta{\~n}eda}, {Cervi{\~n}o}, {Fern{\'a}ndez-Lorenzo}, {Gallego},
  {Gonz{\'a}lez}, {Gonz{\'a}lez-Serrano}, {Nadolny}, {Oteo G{\'o}mez},
  {P{\'e}rez Mart{\'\i}nez}, {Pintos-Castro}, {Povi{\'c}}, \&
  {S{\'a}nchez-Portal}}]{RamonPerez2019b}
{Ram{\'o}n-P{\'e}rez}, M., {Bongiovanni}, {\'A}., {P{\'e}rez Garc{\'\i}a},
  A.~M., {et~al.} 2019{\natexlab{a}}, \aap, 631, A10

\bibitem[{{Ram{\'o}n-P{\'e}rez}
  {et~al.}(2019{\natexlab{b}}){Ram{\'o}n-P{\'e}rez}, {Bongiovanni}, {P{\'e}rez
  Garc{\'\i}a}, {Cepa}, {Nadolny}, {Pintos-Castro}, {Lara-L{\'o}pez}, {Alfaro},
  {Casta{\~n}eda}, {Cervi{\~n}o}, {de Diego}, {Fern{\'a}ndez-Lorenzo},
  {Gallego}, {Gonz{\'a}lez}, {Gonz{\'a}lez-Serrano}, {Oteo G{\'o}mez},
  {P{\'e}rez Mart{\'\i}nez}, {Povi{\'c}}, \&
  {S{\'a}nchez-Portal}}]{RamonPerez2019a}
{Ram{\'o}n-P{\'e}rez}, M., {Bongiovanni}, {\'A}., {P{\'e}rez Garc{\'\i}a},
  A.~M., {et~al.} 2019{\natexlab{b}}, \aap, 631, A11

\bibitem[{{Renzini} \& {Peng}(2015)}]{RP2015}
{Renzini}, A. \& {Peng}, Y.-j. 2015, \apjl, 801, L29

\bibitem[{{Santini} {et~al.}(2015){Santini}, {Ferguson}, {Fontana}, {Mobasher},
  {Barro}, {Castellano}, {Finkelstein}, {Grazian}, {Hsu}, {Lee}, {Lee},
  {Pforr}, {Salvato}, {Wiklind}, {Wuyts}, {Almaini}, {Cooper}, {Galametz},
  {Weiner}, {Amorin}, {Boutsia}, {Conselice}, {Dahlen}, {Dickinson},
  {Giavalisco}, {Grogin}, {Guo}, {Hathi}, {Kocevski}, {Koekemoer},
  {Kurczynski}, {Merlin}, {Mortlock}, {Newman}, {Paris}, {Pentericci},
  {Simons}, \& {Willner}}]{Santini2015}
{Santini}, P., {Ferguson}, H.~C., {Fontana}, A., {et~al.} 2015, \apj, 801, 97

\bibitem[{{Schechter}(1976)}]{Schechter1976}
{Schechter}, P. 1976, \apj, 203, 297

\bibitem[{{Schreiber} {et~al.}(2015){Schreiber}, {Pannella}, {Elbaz},
  {B{\'e}thermin}, {Inami}, {Dickinson}, {Magnelli}, {Wang}, {Aussel}, {Daddi},
  {Juneau}, {Shu}, {Sargent}, {Buat}, {Faber}, {Ferguson}, {Giavalisco},
  {Koekemoer}, {Magdis}, {Morrison}, {Papovich}, {Santini}, \&
  {Scott}}]{Schreiber2015}
{Schreiber}, C., {Pannella}, M., {Elbaz}, D., {et~al.} 2015, \aap, 575, A74

\bibitem[{{Sobral} {et~al.}(2015){Sobral}, {Matthee}, {Best}, {Smail},
  {Khostovan}, {Milvang-Jensen}, {Kim}, {Stott}, {Calhau}, {Nayyeri}, \&
  {Mobasher}}]{Sobral2015}
{Sobral}, D., {Matthee}, J., {Best}, P.~N., {et~al.} 2015, \mnras, 451, 2303

\bibitem[{{Sobral} {et~al.}(2013){Sobral}, {Swinbank}, {Stott}, {Matthee},
  {Bower}, {Smail}, {Best}, {Geach}, \& {Sharples}}]{Sobral2013}
{Sobral}, D., {Swinbank}, A.~M., {Stott}, J.~P., {et~al.} 2013, \apj, 779, 139

\bibitem[{{Somerville} {et~al.}(2004){Somerville}, {Lee}, {Ferguson},
  {Gardner}, {Moustakas}, \& {Giavalisco}}]{somerville04}
{Somerville}, R.~S., {Lee}, K., {Ferguson}, H.~C., {et~al.} 2004, \apjl, 600,
  L171

\bibitem[{{Speagle} {et~al.}(2014){Speagle}, {Steinhardt}, {Capak}, \&
  {Silverman}}]{Speagle2014}
{Speagle}, J.~S., {Steinhardt}, C.~L., {Capak}, P.~L., \& {Silverman}, J.~D.
  2014, \apjs, 214, 15

\bibitem[{{Storey} \& {Hummer}(1995)}]{Storey1995}
{Storey}, P.~J. \& {Hummer}, D.~G. 1995, \mnras, 272, 41

\bibitem[{{Stroe} \& {Sobral}(2015)}]{Stroe15}
{Stroe}, A. \& {Sobral}, D. 2015, \mnras, 453, 242

\bibitem[{{Villar} {et~al.}(2011){Villar}, {Gallego}, {P{\'e}rez-Gonz{\'a}lez},
  {Barro}, {Zamorano}, {Noeske}, \& {Koo}}]{Villar2011}
{Villar}, V., {Gallego}, J., {P{\'e}rez-Gonz{\'a}lez}, P.~G., {et~al.} 2011,
  \apj, 740, 47

\bibitem[{{Villar} {et~al.}(2008){Villar}, {Gallego}, {P{\'e}rez-Gonz{\'a}lez},
  {Pascual}, {Noeske}, {Koo}, {Barro}, \& {Zamorano}}]{Villar2008}
{Villar}, V., {Gallego}, J., {P{\'e}rez-Gonz{\'a}lez}, P.~G., {et~al.} 2008,
  \apj, 677, 169

\bibitem[{{Villaverde} {et~al.}(2010){Villaverde}, {Cervi{\~n}o}, \&
  {Luridiana}}]{Villaverde2010}
{Villaverde}, M., {Cervi{\~n}o}, M., \& {Luridiana}, V. 2010, \aap, 517, A93

\bibitem[{{Whitaker} {et~al.}(2014){Whitaker}, {Franx}, {Leja}, {van Dokkum},
  {Henry}, {Skelton}, {Fumagalli}, {Momcheva}, {Brammer}, {Labb{\'e}},
  {Nelson}, \& {Rigby}}]{Whitaker2014}
{Whitaker}, K.~E., {Franx}, M., {Leja}, J., {et~al.} 2014, \apj, 795, 104

\end{thebibliography}

\onecolumn
\begin{appendix}

\section{Catalogue of \hb{}emitters}
In Table \ref{tab:emitters1} and \ref{tab:emitters2}, we summarise the main properties of our emitters.

\begin{center}
\begin{table}[ht]
\caption[Luminosity functions]{Characteristics of the 27 non-AGN bona fide sources (see Sect. \ref{subsec:selection}).}
\begin{tabular}{ccccccc}
\toprule
ID & $z$ & Flux & EW & $\log M$ & ($g-i$)   &  SFR\\ 
 & & [$\times10^{-17} \mathrm{erg/s/cm}^2$]&[\AA] & [$\mathrm{M}_\odot$]&  &[$\mathrm{M}_\odot\, \mathrm{yr}^{-1}$] \\\midrule
  797                             & 0.863                          & 0.71$\pm0.1$                             & 121.3$_{-23.4}^{+17.3}$                         & 8.60 $\pm$ 0.16                  & 0.99  $\pm$  0.3                       & 0.16$_{-0.03}^{+0.02}$ \\
1223                            & 0.862                          & 0.15$_{-0.05}^{+0.1}$                             & 44.71$_{-18.9}^{+48.1}$                          & 8.50 $\pm$ 0.06                 & 0.54 $\pm$  0.14                       & 0.06$_{-0.02}^{+0.04}$ \\
1981                            & 0.864                          & 0.24$\pm{0.06}$                             & 49.34$_{-13.6}^{+16.23}$                          & 8.27 $\pm$ 0.05                 & 0.17 $\pm$  0.12                       & 0.07$\pm0.02$ \\
2130                            & 0.867                          & 0.44$\pm0.05$                             & 89.5$_{-15.9}^{+21.5}$                          & 7.82 $\pm$ 0.05                      & 0.03 $\pm$  0.14                       & 0.08$\pm0.01$  \\
2236                            & 0.864                          & 1.26$_{-0.1}^{+0.1}$                             & 48.2$_{-5.4}^{+5.9}$                          & 9.01 $\pm$ 0.01                      & 0.38 $\pm$  0.02                       & 0.34$\pm0.03$ \\
2304                            & 0.901                          & 4.59$_{-0.3}^{+0.4}$                              & 14.63$\pm1.09$                          & 10.71 $\pm$ 0.003                      & 1.13 $\pm$  0.006                       & 3.10$\pm0.2$ \\
2447                            & 0.899                          & 0.8$_{-0.08}^{+0.09}$                             & 18.6$_{-2.2}^{+2.5}$                          & 9.31  $\pm$ 0.006                      & 0.49 $\pm$  0.02                       & 0.14$_{-0.01}^{+0.02}$ \\
2623                            & 0.862                          & 0.16$\pm0.06$                             & 24.45$_{-9.4}^{+12.2}$                          & 8.45 $\pm$ 0.02                      & 0.19 $\pm$  0.05                       & 0.04$\pm+0.02$ \\
2644                            & 0.891                          & 0.52$\pm0.06$                            & 41.97$_{-6.2}^{+6.4}$                          & 8.6 $\pm$ 0.01                      & 0.37 $\pm$  0.03                       & 0.09$\pm0.01$ \\
2722                            & 0.893                          & 5.13$_{-0.2}^{+0.3}$                            & 39.3$_{-2.3}^{+2.2}$                          & 9.67 $\pm$ 0.002                      & 0.32 $\pm$  0.005                       & 1.45$\pm0.07$ \\
4971                            & 0.885                          & 0.63$\pm0.1$                             & 37.51$_{-6.74}^{+7.39}$                          & 8.90 $\pm$ 0.01                      & 0.55 $\pm$  0.03                       & 0.15$_{-0.02}^{+0.03}$ \\
5133                            & 0.874                          & 0.66$\pm0.1$                             & 11.40$\pm2.08$                          & 10.15 $\pm$ 0.02                       & 1.31  $\pm$ 0.03                       & 0.3$_{-0.06}^{+0.05}$ \\
5498                            & 0.9                            & 1.84$\pm 0.1$                             & 88.27$_{-8.72}^{+10.28}$                          & 8.93 $\pm$ 0.004                      & 0.06 $\pm$  0.01                       & 0.37$\pm0.3$ \\
5808                            & 0.878                          & 9.23$\pm0.02$                             & 54.74$_{-1.92}^{+1.70}$                          & 9.51 $\pm$ 0.001                      & 0.23 $\pm$  0.002                       & 1.81$_{-0.04}^{+0.05}$ \\
5922                            & 0.884                          & 0.19$_{-0.05}^{+0.06}$                              & 50.16$_{-15.93}^{+22.95}$                           & 8.45 $\pm$ 0.03                      & 0.55 $\pm$  0.07                       & 0.05$\pm0.01$ \\
6474                            & 0.903                          & 0.50$\pm0.07$                             & 55.84$_{-9.98}^{+10.34}$                          & 8.65 $\pm$ 0.01                      & 0.36 $\pm$  0.03                       & 0.09$\pm0.01$ \\
6890                            & 0.862                          & 0.33$\pm0.1$                             & 31.72$_{-9.91}^{+10.30}$                          & 8.46 $\pm$ 0.03                      & 0.35 $\pm$  0.06                       & 0.05$\pm0.02$ \\
7023                            & 0.885                          & 0.84$\pm0.09$                             & 71.01$_{-10.84}^{+12.01}$                          & 8.50 $\pm$ 0.02                      & 0.35  $\pm$ 0.05                       & 0.14$\pm0.2$ \\
7048                            & 0.9                            & 0.75$_{-0.09}^{+0.10}$                             & 19.88$_{-2.87}^{+3}$                          & 9.64 $\pm$ 0.01                       & 0.75 $\pm$  0.02                       & 0.13$\pm0.2$ \\
7467                            & 0.88                           & 0.44$_{-0.05}^{+0.06}$                             & 74.61$_{-13.15}^{+17.47}$                          & 8.56 $\pm$ 0.03                      & 0.56 $\pm$  0.06                       & 0.10$\pm0.01$ \\
7602                            & 0.904                          & 3.42$\pm0.2$                             & 14.68$_{-1.11}^{+1.03}$                          & 10.03 $\pm$ 0.001                      & 0.55 $\pm$  0.003                       & 0.83$\pm0.06$ \\
7629                            & 0.876                          & 2.40$\pm0.2$                           & 36.4$_{-3.27}^{+3.80}$                            & 9.55 $\pm$ 0.004                      & 0.55 $\pm$  0.01                       & 0.56$_{-0.04}^{+0.05}$ \\
7726                            & 0.875                          & 0.41$_{-0.09}^{+0.10}$                             & 46.01$_{-11.53}^{+14.04}$                          & 8.66 $\pm$ 0.02                      & 0.36 $\pm$  0.04                       & 0.07$_{-0.01}^{+0.02}$ \\
7953                            & 0.864                          & 0.50$\pm0.09$                             & 64.97$_{-12.94}^{+15}$                          & 8.39 $\pm$  0.013                      & 0.11 $\pm$  0.04                       & 0.09$\pm0.02$ \\
8010                            & 0.862                          & 0.42$_{-0.08}^{+0.09}$                             & 62.08$_{-16.11}^{+17.68}$                          & 8.82  $\pm$ 0.04                      & 0.55 $\pm$  0.09                       & 0.09 $\pm0.02$ \\
10927                           & 0.861                          & 0.50$_{-0.08}^{+0.07}$                             & 28.90$_{-5.66}^{+4.69}$                          & 8.99 $\pm$ 0.01                      & 0.36 $\pm$  0.01                       & 0.08 $\pm0.01$ \\
11014                           & 0.904                          & 2.68$\pm0.2$                             & 17.35$_{-1.2}^{+1.1}$                          & 10.26 $\pm$ 0.002                      & 0.94 $\pm$  0.004                       & 1.29$\pm0.08$ \\
\hline
\end{tabular}

\label{tab:emitters1}
\end{table}
\end{center}

\begin{table}[ht]
\caption[Luminosity functions]{Same as Table \ref{tab:emitters1}, but for the 13 sources which cannot been considered as bona fide ones (see Sect. \ref{subsec:selection}.).}
\begin{tabular}{ccccccc}
\toprule
ID & $z$ & Flux & EW & $\log M$ & ($g-i$)   &  SFR\\ 
 & & [$\times10^{-17} \mathrm{erg/s/cm}^2$]&[\AA] & [$\mathrm{M}_\odot$]&  &[$\mathrm{M}_\odot\, \mathrm{yr}^{-1}$] \\\midrule
2755                            & 0.862                          & 0.52$\pm0.13$                              & 47.5$_{-12.9}^{+14.3}$                          & 8.22 $\pm$ 0.16                      & 0.81 $\pm$  0.3                         & 0.56$\pm0.14$ \\
4460                            & 0.855                          & 0.41$_{-0.13}^{+0.22}$                             & 72.90$_{-22.7}^{+48.9}$                          & 10.09 $\pm$ 0.16                       & 1.24  $\pm$   0.24                       & 0.08$_{-0.02}^{+0.04}$ \\
5156                            & 0.873                          & 0.46$\pm 0.09$                             & 30.18$_{-6.97}^{+6.82}$                          & 8.78 $\pm$ 0.02                      & 0.55 $\pm$  0.04                       & 0.11$\pm0.21$ \\
5864                            & 0.873                          & 0.24$\pm0.06$                             & 39.58$_{-10.83}^{+13.13}$                          & 8.32 $\pm$ 0.02                      & 0.36 $\pm$  0.06                       & 0.04 $\pm0.01$  \\
6456                            & 0.903                          & 3.90$\pm0.2$                             & 56.37$_{-2.93}^{+2.83}$                           & 9.35 $\pm$ 0.001                      & 0.09 $\pm$  0.004                       & 0.80$\pm0.03$ \\
6838                            & 0.862                          & 0.12$\pm0.03$                             & 81.34$_{-28.31}^{+33.87}$                          & 7.78 $\pm$ 0.001                        & 0.35                          & 0.03$\pm0.01$ \\
7868                            & 0.886                          & 0.78$_{-0.09}^{+0.08}$                             & 142.97$_{-27.89}^{+34.03}$                         & 8.35 $\pm$ 0.02                      & 0.14 $\pm$  0.04                       & 0.16$\pm0.02$ \\
8187                            & 0.892                          & 1.48$_{-0.09}^{+0.08}$                             & 80.12$_{-7.36}^{+7.20}$                          & 9.03 $\pm$ 0.01                      & 0.54 $\pm$  0.02                       & 0.35$\pm0.02$  \\
9344                            & 0.862                          & 0.19$_{-0.06}^{+0.07}$                             & 37.39$_{-14.33}^{+15.22}$                          & 7.60 $\pm$ 0.1                       & 0.25 $\pm$  0.15                       & 0.07$\pm0.03$ \\
9927                            & 0.861                          & 0.17$_{-0.06}^{+0.07}$                             & 39.49$_{-13.55}^{+20.30}$                          & 7.97 $\pm$ 0.02                      & 0.21 $\pm$  0.06                        & 0.05$\pm0.02$  \\
10097                           & 0.86                           & 0.25$\pm0.05$                             & 77.41$_{-18.58}^{+23.22}$                           & 7.82 $\pm$ 0.3                       & 0.34   $\pm$  0.91                       & 0.04$\pm0.01$ \\
10988                           & 0.863                          & 0.19$_{-0.06}^{+0.07}$                             & 52.17$_{-20.20}^{+31.32}$                           & 7.91 $\pm$ 0.03                      & 0.19 $\pm$  0.07                       & 0.05$\pm0.02$  \\
11063                           & 0.864                          & 0.20$\pm0.06$                             & 41.76$_{-13.2}^{+16.2}$                          & 8.37 $\pm$ 0.02                      & 0.31 $\pm$  0.05                       & 0.05$\pm0.02$  \\ \hline
\end{tabular}

\label{tab:emitters2}
\end{table}

\end{appendix}
\end{document}